%% file: Gaidos.tex
\documentclass[fleqn,usenatbib]{mnras}

\include{newcommand_gen}

\usepackage{newtxtext,newtxmath,lineno}

\usepackage[T1]{fontenc}
\usepackage{ae,aecompl}

\usepackage{graphicx}	
\usepackage{amsmath}	
\usepackage{amssymb}	

\title[]{Zodiacal Exoplanets in Time. X. The Orbit and Atmosphere of the Young "Neptune Desert"-Dwelling Planet K2-100b}

\author[Gaidos et al.]{
E. Gaidos\thanks{E-mail: gaidos@hawaii.edu}$^{1,2,3}$, T. Hirano$^{4}$, A. W. Mann$^{5}$, D. A. Owens$^{5}$, T. A. Berger$^{6}$,
K. France$^{7}$, 
\newauthor
A. Vanderburg$^{8}$,
H. Harakawa$^{9,10}$, 
K. W. Hodapp$^{11}$, 
M. Ishizuka$^{12}$, 
S. Jacobson$^{11}$,
\newauthor
M. Konishi$^{13}$, 
T. Kotani$^{10,14,15}$,
T. Kudo$^{9,10}$, 
T. Kurokawa$^{14,16}$,
M. Kuzuhara$^{10,14}$, 
\newauthor
J. Nishikawa$^{10,14}$,
M. Omiya$^{10,14}$,
T. Serizawa$^{16}$,
M. Tamura$^{10,12,14}$ 
\& A. Ueda$^{14}$\\
$^{1}$Department of Earth Sciences, University of Hawai'i at M\={a}noa, Honolulu, HI  96822, USA\\
$^{2}$Kavli Institute for Theoretical Physics, UC Santa Barbara, Santa Barbara, CA 93106\\
$^{3}$Institute for Astrophysics, Georg-August-Universit\"{a}t G\"{o}ttingen, 37077 G\"{o}ttingen, Germany\\
$^{4}$Department of Earth \& Planetary Sciences, Tokyo Institute of Technology, 2-12-1 Ookayama, Meguro-ku, Tokyo 152-8551, Japan\\
$^{5}$Deptartment of Physics \& Astronomy, University of North Carolina at Chapel Hill, Chapel Hill, NC, USA\\
$^{6}$Institute for Astronomy, University of Hawaii at M\={a}noa, Honolulu, HI 96822, USA\\
$^{7}$Laboratory for Atmospheric and Space Physics, University of Colorado, Boulder, CO 80309, USA\\
$^{8}$Department of Astronomy, The University of Texas at Austin, Austin, TX 78712, USA\\
$^{9}$Subaru Telescope, 650 N. Aohoku Place, Hilo, HI 96720, USA\\
$^{10}$Astrobiology Center, NINS, 2-21-1 Osawa, Mitaka, Tokyo 181-8588, Japan\\
$^{11}$University of Hawaii, Institute for Astronomy, 640 N. Aohoku Place, Hilo, HI 96720, USA\\
$^{12}$Department of Astronomy, Graduate School of Science, The University of Tokyo, 7-3-1 Hongo, Bunkyo-ku, Tokyo 113-0033, Japan\\
$^{13}$Faculty of Science and Technology, Oita University, 700 Dannoharu, Oita 870-1192, Japan\\
$^{14}$National Astronomical Observatory of Japan, NINS, 2-21-1 Osawa, Mitaka, Tokyo 181-8588, Japan\\
$^{15}$Department of Astronomy, School of Science, The Graduate University for Advanced Studies (SOKENDAI), 2-21-1 Osawa, Mitaka, Tokyo, Japan\\
$^{16}$Tokyo University of Agriculture and Technology, 3-8-1, Saiwai-cho, Fuchu, Tokyo, 183-0054, Japan\\
}

\date{Accepted 2020 March 26. Received 2020 March 22; in original form 2020 February 7}

\pubyear{2020}

\begin{document}
\label{firstpage}
\pagerange{\pageref{firstpage}--\pageref{lastpage}}
\maketitle

\begin{abstract}
We obtained high-resolution infrared spectroscopy and short-cadence photometry of the 600-800 Myr Praesepe star K2-100 during transits of its 1.67-day planet.  This Neptune-size object, discovered by the NASA \ktwo{} mission, is an interloper in the ``desert" of planets with similar radii on short period orbits. Our observations can be used to understand its origin and evolution by constraining the orbital eccentricity by transit fitting, measuring the spin-orbit obliquity by the Rossiter-McLaughlin effect, and detecting any extended, escaping hydrogen-helium envelope with the 10830 \AA{} line of neutral helium in the 2s$^3$S triplet state.  Transit photometry with 1-min cadence was obtained by the \ktwo{} satellite during Campaign 18 and transit spectra were obtained with the IRD spectrograph on the Subaru telescope.  While the elevated activity of K2-100 prevented us from detecting the Rossiter-McLaughlin effect, the new photometry combined with revised stellar parameters allowed us to constrain the eccentricity to $e < 0.15/0.28$ with 90\%/99\% confidence.  We modeled atmospheric escape as an isothermal, spherically symmetric Parker wind, with photochemistry driven by UV radiation that we estimate by combining the observed spectrum of the active Sun with calibrations from observations of K2-100 and similar young stars in the nearby Hyades cluster.  Our non-detection ($<5.7$m\AA) of a transit-associated \hei{} line limits mass loss of a solar-composition atmosphere through a $T\le10000$K wind to $<0.3$ \mearth{}~Gyr$^{-1}$.  Either K2-100b is an exceptional desert-dwelling planet, or its mass loss is occurring at a lower rate over a longer interval, consistent with a core accretion-powered scenario for escape.
\end{abstract}




\section{Introduction}
\label{sec:intro}

\begin{figure}
	\includegraphics[width=\columnwidth]{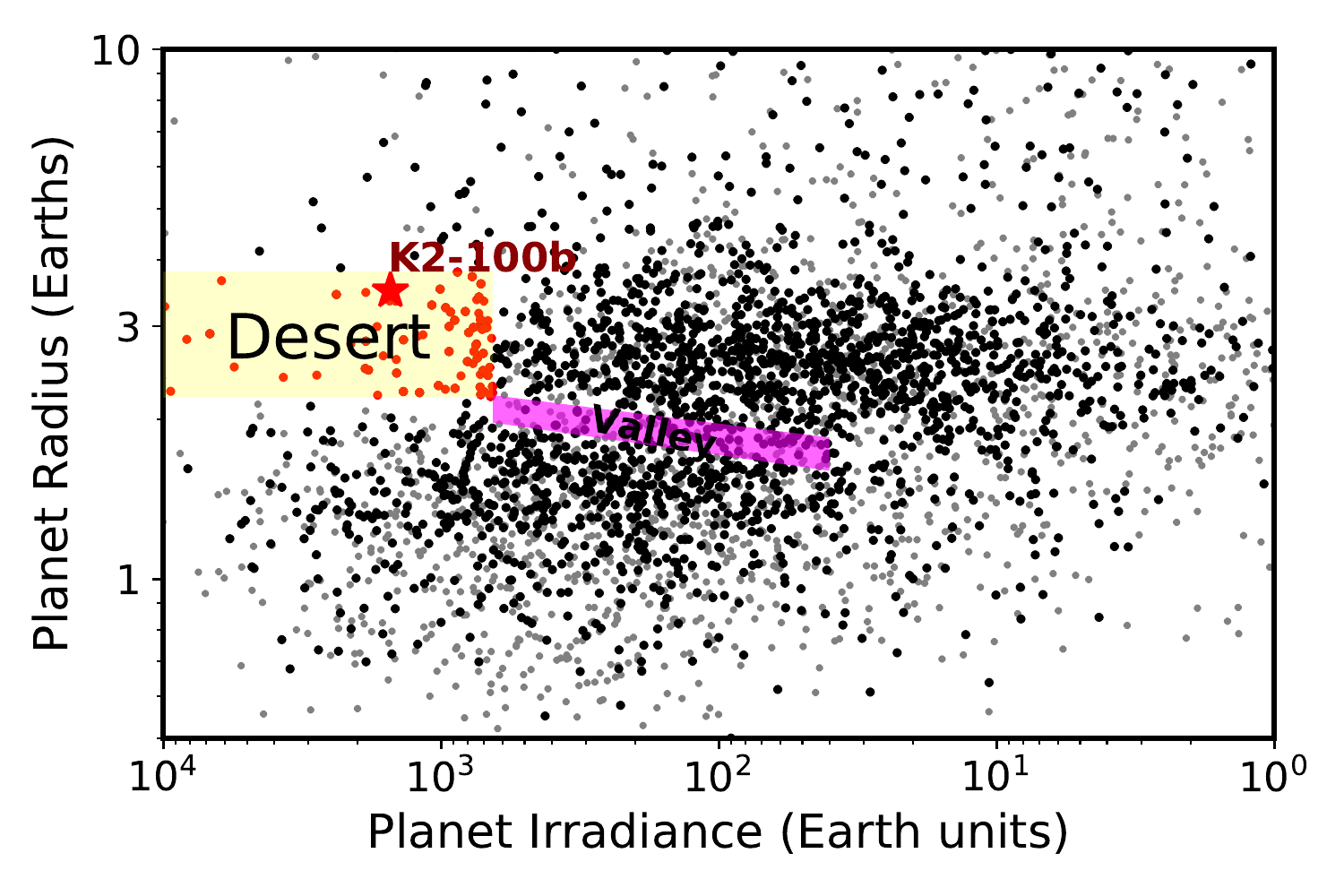}
    \caption{Radius and irradiance of \kepler-detected planets, based on \citet{Berger2018}, showing the location of K2-100b in the ``Neptune desert".}  
    \label{fig:rad-irrad}
\end{figure}

The \kepler\ mission showed that planets are common around other stars and revealed structure in the distribution of planets with size and irradiance (or orbital separation) from their host stars (Fig. \ref{fig:rad-irrad}):  This is an important clue to how planets formed, migrated, and evolved.  One such feature is the paucity of planets with radii of 2-4\rearth{} on close-in (P $\lesssim$3~day) orbits.  This region is known as the ``Neptune desert" \citep{Lundkvist2016,Berger2018}, since measured masses and radii of planets in that radius range indicate a hydrogen-helium gas-rich composition analogous to Neptune \citep{Weiss2014,Rogers2015}.  It has been proposed that the "desert" is the product of the loss of H-He atmospheres by core accretion-powered escape \citep{Ginzburg2016}, X-ray/UV (XUV)-powered (photoevaporative) escape \citep[e.g.,][]{Owen2018b}, or, if the planets accreted in situ \citep{Chiang2013}, inhibition of accretion of such envelopes by the proximity of the host star, or lack of gas in the inner protoplanetary disk.     

Recently, interlopers in the desert have been identified \citep[e.g.,][]{Berger2018,West2019,Hobson2019}.  
There are several possible explanations for this population:  Planets could move into the desert as a result of the central star evolving from the main sequence and increasing in luminosity; many of the exceptions found by \citet{Berger2018} fall in this category.  Some planets may be much more massive, with high molecular-weight envelopes that resist escape to space.  They could be the result of catastrophic escape of the H-He envelopes of close-in giant planets ("hot Jupiters") that have shrunk to Neptune size \citep{Dong2018}.  The mass of these planets can be measured by precision radial velocity measurements \citep[e.g.,][]{Espinoza2016}.  Finally, in young planetary systems, escape of the H-He envelope may be incomplete and still ongoing.  Since an envelope of at least 5\% by mass is required to produce a Neptune-like object \citep{Ginzburg2016}, and the time scale for escape is supposed to be similar to that of decreasing stellar magnetic activity (a few hundred Myr), one expects an escape rate of at least 1 \mearthpergyr{} during that epoch.

The last two possibilities can be tested with observations that constrain the orbit and atmosphere of such a planet.  Some hot Jupiter orbits are inclined with respect to the rotational axis of the star, and thus also the presumed plane of the natal disk.  Such orbits can arise when the planet is originally driven onto a highly inclined and eccentric orbit by planet-planet scattering, or a Kozai resonance with an outer stellar companion before circularization by the planetary tide \citep[][and references therein]{Dawson2018}.  Sufficiently rapid escape of most of the mass of the giant planet would halt the process of circularization,  leaving a Neptune-mass planet on an inclined, moderately eccentric orbit.  Orbital inclination or, in the case of a transiting planet, stellar obliquity, can be derived from the Rossiter-McLaughlin effect -- the change in a star's apparent RV due to the partial occultation of the rotating stellar disk by a transiting planet \citep{Triaud2017}.  In the case of rapidly rotating stars it is sometimes possible to perform a detailed analysis of the changes in the stellar line shape, a method often referred to as ``Doppler tomography" \citep{Albrecht2007,CollierCameron2010}.  While large orbital eccentricities of giant planets are readily revealed by RV measurements, modest orbital eccentricities of smaller planets are more difficult to detect.  One approach for transiting planets is to compare the duration of a transit to that expected from a stellar density established by other observations and an inferred transit impact parameter \citep[e.g.,][]{VanEylen2015,Xie2016,Mann2017b,VanEylen2019}.

Around young stars, ongoing loss of an atmosphere should manifest itself as an extended, escaping atmosphere that could be detected by spectroscopy during transit.  Escaping neutral hydrogen (H\,I) has been detected by observations in the line of Lyman $\alpha$ \citep[e.g.,][]{VidalMadjar2003,Lecavelier2010,Kulow2014,Ehrenreich2015} but this can be done only from space and is impeded by the absorption of the line core by interstellar H\,I, interfering geocoronal Lyman $\alpha$ emission, and variable stellar chromosphere emission.  Neutral helium (\hei) in its metastable 2s$^{3}$S or ``triplet" state can be detected via absorption in a near-infrared set of lines around 10830\AA{} that are accessible from the ground \citep{Seager2000,Oklopovcic2018}.  The first detection was with HST from space \citep{Spake2018} but the introduction of high-resolution infrared spectrographs at facilities on the ground has led to detections or claimed detection of \hei{} among several transiting planets \citep{Allart2018,Mansfield2018,Nortmann2018,Salz2018,Allart2019,Ninan2019} and some non-detections \citep{Moutou2003,Nortmann2018,Kreidberg2018,Crossfield2019}.  

K2-100b is a 3.5\rearth{} planet that was detected by the \emph{K2} mission on a 1.67-day orbit transiting the G0-type star K2-100 (TYC 1398-142-1 or EPIC 211990866) \citep{Mann2017}, and which occupies the Neptune desert (Fig. \ref{fig:rad-irrad}).   The star is a member of the nearby \citep[189 pc,][]{Gaia2018b} young open cluster M-44, the Beehive or Praesepe.  The age of the cluster is not known precisely; it is variously estimated as 625 Myr \citep{Perrymann1998}, 850 Myr \citep{Brandt2015}, and 590 Myr \citep{Gossage2018,Schroeder2019}, with an uncertainty of about 100 Myr.  The planet's orbital ephemeris were refined with ground-based detections by \citet{Stefansson2018}, allowing for transits to be predicted with accuracy of a minute or less.  \citet{Mann2017} pointed out that the transit duration of ``b" and the estimated density of the star are suggestive of an eccentric orbit ($e = 0.24 \pm^{0.19}_{0.12}$), depending on the impact parameter.  \citet{Stefansson2018} also derived a transit-based density that was $\sim3\times$ higher than the one based on multi-band photometry and a \emph{Hipparcos} parallax of the cluster \citep{Mann2017}, consistent with an eccentric orbit.  However, this is based on transit photometry obtained with a 30-min cadence, only one-third of the transit duration.  K2-100b's host star is highly rotationally variable, challenging both transit and RV measurements.  \citet{Barragan2019} used a Gaussian process-regression of RV measurements with spectroscopic indicators of stellar activity to infer a mass of $22 \pm 6$\mearth{}, i.e. at 3$\sigma$ significance.  The mass and radius combination point to a Neptune-like composition and thus a significant H-He envelope which is normally not associated with planets this close to their host star.  To constrain the orbit and atmosphere of K2-100b, we analyzed short-cadence photometry obtained by \ktwo{} during Campaign 18 (2018), and high-resolution infrared spectra obtained with the IRD spectrograph on the Subaru telescope during a transit in 2019. 

\section{Observations and Data Reduction}
\label{sec:observations}

\subsection{\ktwo{} and literature photometry}
\label{sec:ktwo}
\ktwo\ observed K2-100 as part of Campaign 5 (April 27, 2015 to July 10, 2015) in long-cadence mode (30 min) and again during Campaign 18 (May 12, 2018 to July 2, 2018) in short-cadence mode (1 min). The Campaign 5 observations were used to identify the planet, and are described in more detail in the discovery paper \citep{Mann2017}. The Campaign 18 short-cadence observations were proposed to improve the parameters of K2-100b and search for long-period planets (GO18006, GO18023, GO18027, and GO18036), and were described previously in \citet{Barragan2019}. 

The \ktwo\ light curves exhibit variations due both to telescope drift \citep{VanCleve2016} as well as astrophysical variability from stellar rotation and transits. We fit for all these effects simultaneously as described in detail in \citet{Becker2015} and \citet{Vanderburg2016}. The simultaneous fit is required to avoid changing the transit results as the stellar variation/flat field fit is biased by the presence of variability that cannot be easily explained by the model \citep[e.g., ][]{Grunblatt:2016aa}. Stellar flares still present in the processed light curve are manually identified and removed. The resulting light curve was used in our Markov Chain Monte Carlo (MCMC) analysis (Section~\ref{sec:transit}). 

In addition to the \ktwo\ data, four transits of K2-100b were observed from the ground. The first used the Engineered Diffuser with the Astrophysical Research Council Telescope Imaging Camera (ARCTIC) imager on the ARC 3.5 m Telescope at ApachePoint Observatory. The other three used the MuSCAT2 multi-bandpass photometer installed in the Carlos Sanchez Telescope (TCS) in the Teide observatory.   The diffuser transit observations are described in \citet{Stefansson2018}, while the MuSCAT2 observations are described in \citet{Barragan2019}. For all data, we used the provided light curve for our MCMC analysis, only normalizing the out-of-transit baseline and updating the uncertainties based on the estimates in \citet{Barragan2019}.

\subsection{Infrared spectroscopy}
\label{sec:ird}

Spectra of K2-100 were obtained during the transit of ``b" on 24 March 2019 with the IRD infrared echelle spectrograph \citep{Kotani2018} on the Subaru telescope on Maunakea.  IRD accepts light from the target via a multi-mode fiber with input aperture of $0\farcs48$ fed by the facility adaptive optics; a second fiber provides light from the laser comb wavelength reference source.   IRD covers the $Y$-, $J$-, and $H$-bands simultaneously with a spectral resolution of 70,000.  Using the ephemeris of \citet{Stefansson2018}, the transit mid-point occurred at $\mathrm{JD}=245866.881$ (UT 9:09:00) with a prediction accuracy of 50~sec.    Observations commenced at $\mathrm{JD}=2458566.81$ and ended at $\mathrm{JD}=2458566.98$ and consisted of a series of 12-min integrations, the first starting 58 minutes before the predicted ingress (UT 8:21) and the last completing 93 minutes after the egress (UT 9:58).  (The transit duration is 97 minutes).   Sky conditions were scattered high clouds; the attenuation recorded by the CFHT Skyprobe camera \citep{Steinbring2009} shows variable attenuation of up to 2 magnitudes, but generally less than 1.5 magnitudes, during all but the second hour of the night.  This resulted in SNR that was 50-60\% of expected. To correct for telluric absorption, we observed two A0-type telluric standard stars (HIP 22923 and HD 71906) immediately before observing K2-100, but these spectra are essentially flat in the vicinity of the \hei{} line, and due to the large variation in airmass and water content over the night, plus the fact that the transit measurement is differential, we ultimately did not make this correction.

Using the \texttt{IRAF} echelle package \citep{Tody1986, Tody1993} and our own custom software, we applied bias subtraction, flat-fielding, and scattered-light subtraction to the spectral images before extracting one-dimensional spectra.  A wavelength solution was first determined based on the emission lines of a Th-Ar comparison-lamp, then refined using the spectrum of IRD's laser comb.  We measured the relative Doppler shift of each spectrum and obtained a radial velocity (RV) using our customized code (Hirano et al. submitted).  Briefly, the pipeline obtains RVs for individual frames by forward modeling of the observed spectrum relative to a template spectrum of the star.  The stellar template is produced by removing telluric lines using rapidly-rotating nearly featureless spectrum of an A0 star, and deconvolving the instantaneous point response function of the spectrograph.  For our observations of K2-100, the pipeline yielded RVs with a precision of $13-29$ m s$^{-1}$, limited by the relatively low SNR and our ability to remove telluric contamination. 

\section{Analysis}
\label{sec:analysis}

\subsection{Stellar Parameters}
\label{sec:parameters}

\citet{Mann2017} derived a stellar effective temperature \teff{} of $6120 \pm 90$K using the color-temperature relations of \citet{Ramirez2005} and \citet{Pinsonneault2012}.  This estimate agrees with the value of \teff=$6180$ and $\log g =4.51$ independently derived by 
\citet{Petigura2018b} matching empirical templates to a Keck HIRES spectrum.  \citet{Sousa2018} derived a slightly cooler \teff{} of $5981\pm44$K, $\log g = 4.32\pm0.15$, and [Fe/H] = $+0.18\pm0.04$ (a metallicity consistent with the Praesepe mean) by a line-by-line analysis on a VLT-UVES echelle spectrum. 

We re-compute the stellar radius using the \gaia\ DR2 parallax of $5.26 \pm 0.07$ mas, which corresponds to a bias-corrected distance of $189 \pm 2.5$~pc.  Using the temperature and bolometric flux derived by \citet{Mann2017} we estimate a luminosity of $1.93\pm 0.05$\lsun\ and a radius of $1.23 \pm 0.04$\rsun.  Lack of companions is indicated by absence of resolved sources in AO imaging \citep{Mann2017}, no drift in RV \citep{Barragan2019}, and a value of 1.04 for the \gaia\ Renormalized Unit Weight Error (RUWE), a measure of the deviation of the astrometry from a single-star solution \citep{Kervella2019}.  We retain the extinction correction of \citet{Mann2017} since the new reddening maps of \citet{Green2019} provides no additional information.

We utilized {\tt isoclassify} \citep{Huber2017} with the logarithmic-linear grid detailed in \citet{Berger2020} to perform an isochrone analysis of K2-100. We placed a $790 \pm 30$ Myr prior on its age to ensure our derived age matches cluster estimates. We found a stellar radius of $1.20 \pm 0.02$\rsun{}, a stellar mass of $1.23 \pm 0.01 $\msun, and gravity $\log g = 4.37 \pm 0.01$, consistent with previous estimates \citep{Mann2017}.  Combined with the precisely established orbital period this gives a semi-major axis for ``b" of $0.0296 \pm 0.0005$ AU.  

\subsection{Transit lightcurve and orbit}
\label{sec:transit}

With the revised mass and radius of K2-100 we re-visit the question of the orbit of the planet.  To analyze the transits for K2-100b, we used the {\tt MISTTBORN} (MCMC Interface for Synthesis of Transits, Tomography, Binaries, and Others of a Relevant Nature) code\footnote{\href{https://github.com/captain-exoplanet/misttborn}{https://github.com/captain-exoplanet/misttborn}}, first described in \citet{Mann2016a}. {\tt MISTTBORN} combines the {\tt emcee} Python module \citep{Foreman-Mackey2013} for MCMC and the {\tt batman} package \citep{Kreidberg2015}, which uses the \citet{MandelAgol2002} model of a transiting exoplanet.  The free parameters are planet-to-star radius ratio ($R_{p}/R_{*}$), impact parameter (\textit{b}), orbital period (\textit{P}), epoch of the first transit midpoint (\textit{$T_{0}$}), two parameters that describe the eccentricity and argument of periastron ($\sqrt{e} \sin \omega$ and $\sqrt{e} \cos \omega$), mean stellar density ($\rho_{*}$), and two limb-darkening parameters for each of four wavelengths ({\textit Kepler}, SDSS $r'$, SDSS $i'$ and SDSS $z'$). For limb-darkening, we used the triangular limb-darkening parameters ($q_1$ and $q_2$) described by \citet{Kipping2013} to uniformly explore the physically allowed region of parameter space. We performed one MCMC fit with the parameters above and a Gaussian prior on $\rho_*$ from our derived stellar parameters, and a second fit locking $e$ and $\omega$ at 0 and with a uniform prior on $\rho_*$.  We placed Gaussian priors on each limb-darkening parameter derived from the LDTK toolkit \citep{2015MNRAS.453.3821P}, which uses the \citet{2013A&A...553A...6H} stellar atmosphere models. The formal errors on the derived coefficients are generally 0.01-0.05, but we adopt broader priors (0.05-0.1) to account for differences between model predictions.  We ran our MCMC using 150 walkers, each for 90,000 steps following a burn-in of 10,000 steps. An examination of the autocorrelation time confirmed this was more than sufficient for convergence. 

\begin{figure}
	\includegraphics[width=\columnwidth]{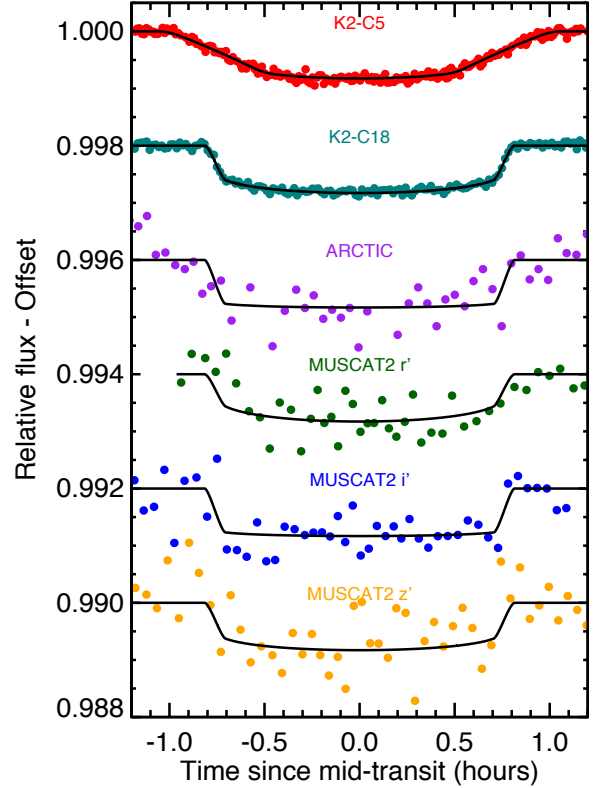}
    \caption{Phase-folded light curves of transits of K2-100b from \ktwo{} Campaigns 5 and 18, ARCTIC, and MuSCAT2 observations, with the MuSCAT2 data separated by pass-band. For clarity, the data from \ktwo\ C18, ARCTIC, and MUSCAT, are binned by 40, 10, and 20 measurements, respectively. The best-fit models are shown in black. Campaign 5 data from \ktwo\ was obtained with 30 min cadence, which smooths out the transit ingress and egress.}  
    \label{fig:lc}
\end{figure}

\begin{figure}
	\includegraphics[width=\columnwidth]{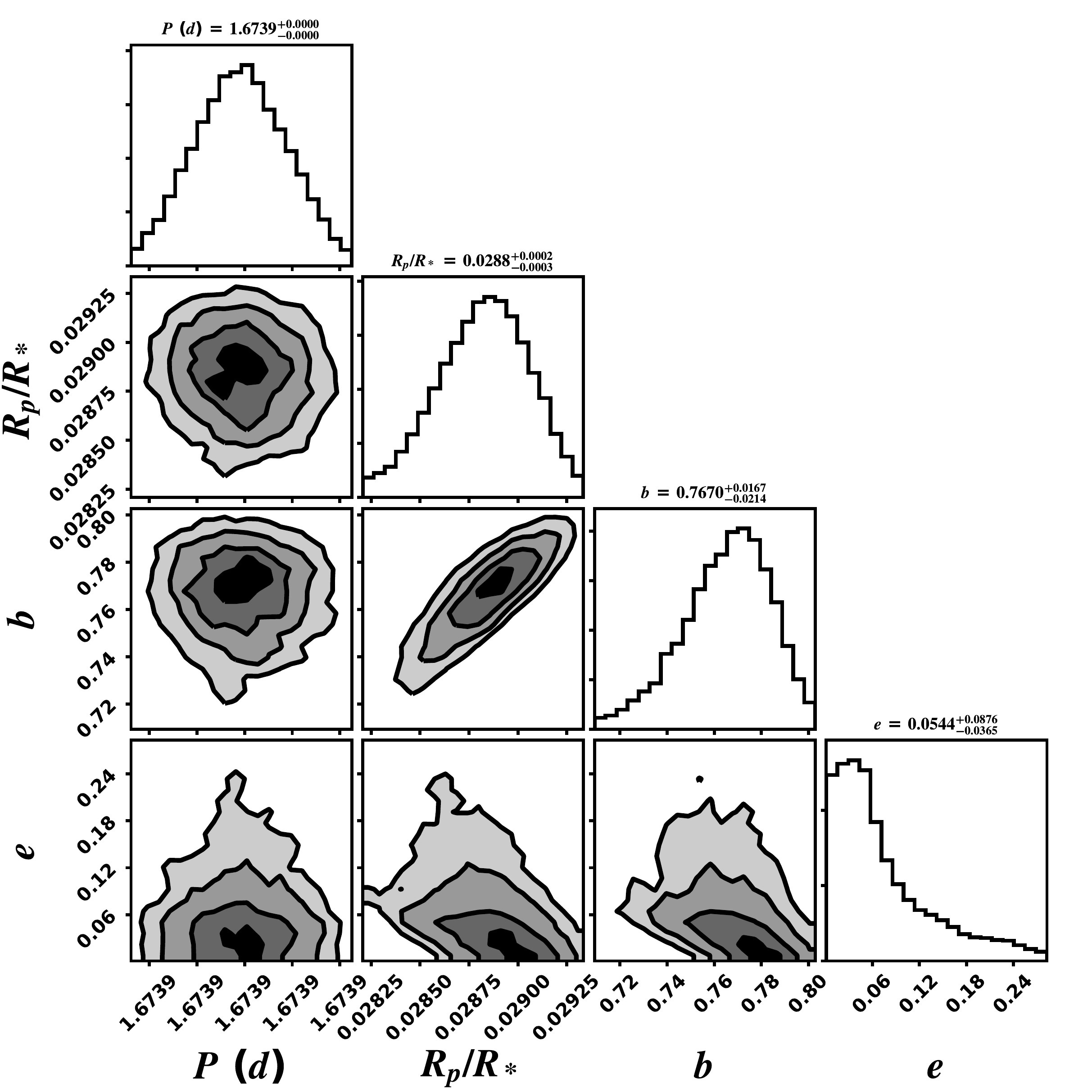}
    \caption{Corner plot of the posteriors of the fitting parameters in our MCMC analysis of the K2-100 light curves, made with {\tt corner.py} \citep{corner}. Contours show 1$\sigma$, 2$\sigma$, and 3$\sigma$ of the 2-dimensional distributions. Eccentricity ($e$) is not directly fit; instead we show the $e$ value derived from $\sqrt{e} \sin \omega$ and $\sqrt{e} \cos \omega$. A small number ($\simeq2\%$) of points fall outside of the plotted parameter spaces.
    }  
    \label{fig:params}
\end{figure}

Results of each fit are shown in Table~\ref{tab:param}, the model light curves with the best-fit parameters (highest likelihood) for each dataset are plotted in Fig. \ref{fig:lc}, and correlations between the major parameters are shown in Fig. \ref{fig:params}. Our results are broadly consistent with the most recent lightcurve analysis from \citet{Barragan2019}. This includes a preference for low or zero orbital eccentricity ($e<0.15$ with 90\% confidence, and $e<0.28$ at 99\% confidence) compared to earlier analysis lacking short-cadence \ktwo\ data that suggested a non-zero orbital eccentricity \citep[$e = 0.28 \pm^{0.19}_{0.12}$][]{Mann2017}.

\subsection{Stellar rotation}
\label{sec:rotation}
\begin{figure}
	\includegraphics[width=\columnwidth]{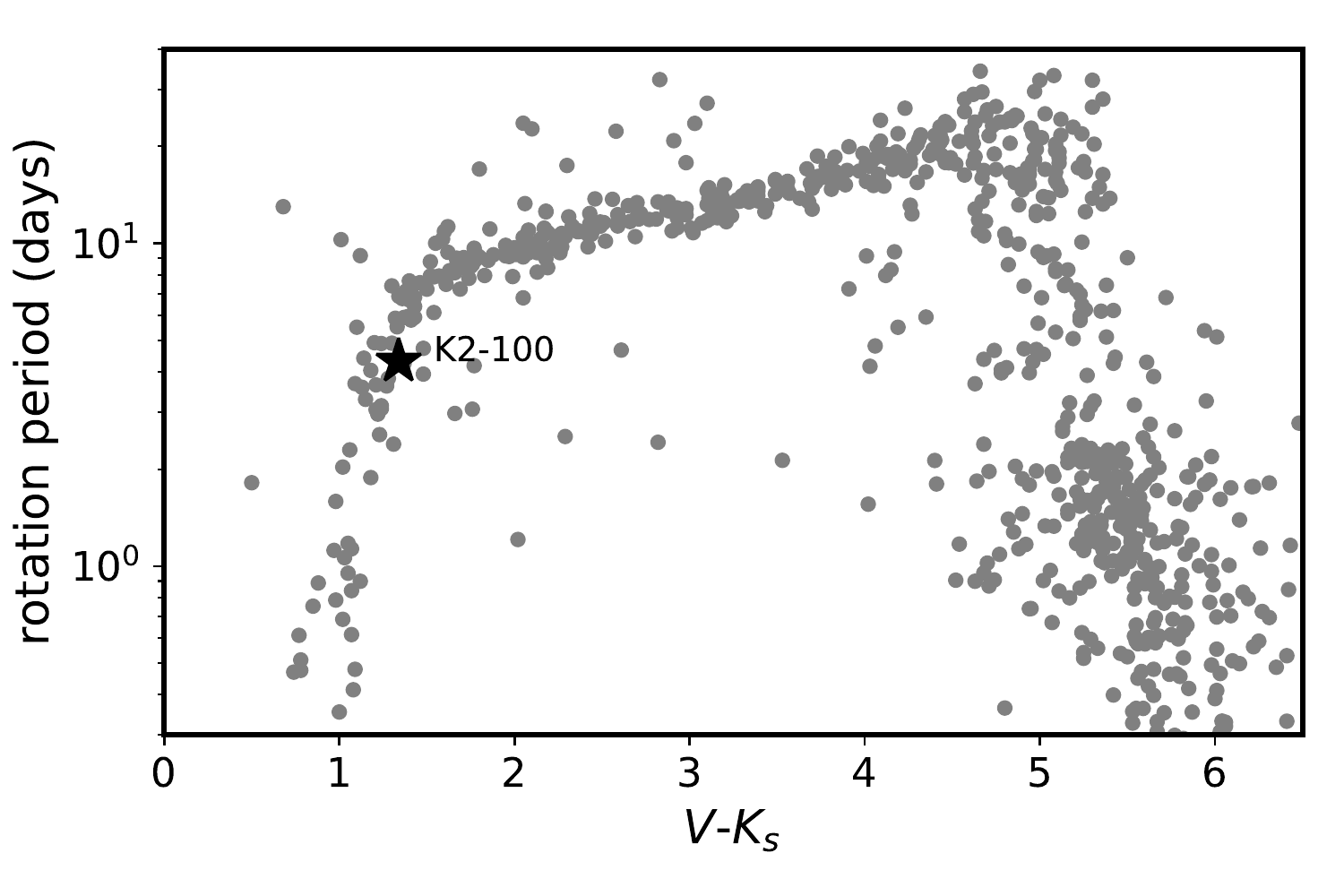}
    \caption{Rotation period of Praesepe stars vs. $V-K_s$ from \citet{Rebull2017}.  The position of K2-100 ($P=4.315$d) is marked as a star symbol.}  
    \label{fig:rotation}
\end{figure}

By analyzing time-series of spectroscopic activity signals, \citet{Barragan2019} refined the rotation period of K2-100 to 4.315 days.  The star lies within the locus of Praesepe stars in a period-color diagram \citep[Fig. \ref{fig:rotation},][]{Rebull2017} and at the upper boundary of the Kraft break where the convective zone is thin and magnetic braking is weak \citep{Kraft1967}.  Nevertheless, a comparison between the 125 Myr-old Pleiades cluster and Praesepe shows significant spin-down among stars of this $V-K_s$ color in the intervening $\sim$0.5 Myr \citep{Rebull2017}, thus magnetic activity and braking must be ongoing, as is evidenced by the star's rotational variability.  There is no indication that the existence of the planet has influenced the star's rotational history.  A constraint on $v \sin i$ was derived using the cross-correlation function for IRD spectra and comparing to model spectra with varying amount of rotational broadening (see Fig. \ref{fig:ccf}). Details are given in Appendix \ref{sec:appendix}. 
The resulting value, $14.4 \pm 0.5$ \kms, is comparable to the value of $14.8 \pm 0.8$ \kms{} derived by  \citet{Mann2017} using near-infrared echelle spectra (IGRINS).  \cite{Barragan2019} estimated $14 \pm 2$ \kms{} using optical echelle (HARPS) data,  \citet{Petigura2018b} derived $v \sin i = 13.6 \pm 1$ \kms{} based on matching a HIRES spectrum to templates, and \citet{Sousa2018} derived $v \sin i = 13.6 \pm 0.3$ km/s using a UVES spectrum and spectral synthesis.   We combine these five completely independent measurements into a single estimate:  $13.9 \pm 0.3$\kms{}.  The expected microturbulence parameter for this star based on its \teff\ (6120 K) and \logg\ (4.36) from \citet{Mann2017} is 1.23 \kms\ based on the calibrated relation of \citet{Bruntt2012}, and is ignored in all calculations here.  Macroturbulence is expected to be $\approx 4.5$ \kms{} \citep{Doyle2014} and \citet{Sousa2018} find 4.19 \kms, but this will primarily affect the wings of strong lines and is also not included here.   The derived rotation velocity based on the rotation period (4.315 days) and $R_*$ is $14.4 \pm 0.5$ \kms\, equal to within $1\sigma$ of our estimate for $v \sin i$.  We conclude the stellar rotational inclination is close to 90 deg and at the 1$\sigma$ confidence level $>70$ deg.  

\begin{figure}
	\includegraphics[width=\columnwidth]{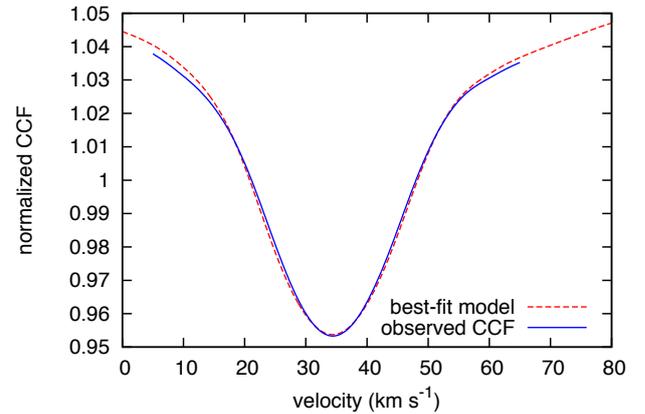}
    \caption{Cross-correlation function of the IRD spectrum of K2-100 and a comparison to the best-fit model (red dashed line) with $v \sin i = 14.4$ \kms.}  
    \label{fig:ccf}
\end{figure}

\subsection{Rossiter-McLaughlin effect and Doppler tomography}
\label{sec:rm}
\begin{figure}
	\includegraphics[width=\columnwidth]{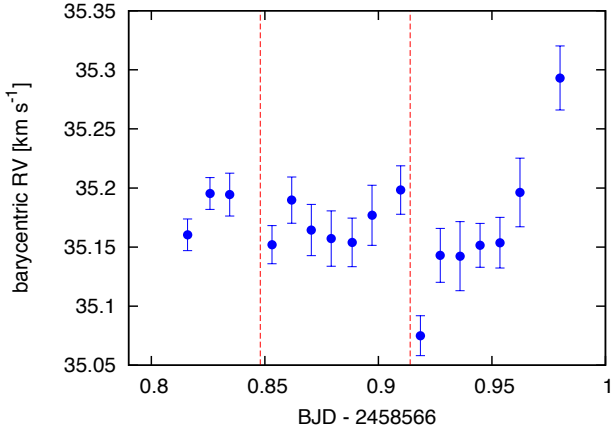}
    \caption{Barycentric radial velocities of K2-100 obtained by analysis of IRD spectra. The transit ingress and egress times are marked by the vertical red dashed lines. }  
    \label{fig:rvs}
\end{figure}

The expected amplitude of the RV anomaly due to the Rossiter-McLaughlin effect is 
\begin{eqnarray}
\Delta v_\mathrm{R-M} \approx 0.7 \delta \cdot v\sin i\cdot \sqrt{1-b^2},
\end{eqnarray}
where $\delta$ and $b$ are the transit depth and impact parameter, respectively.  The 0.7 factor is a crude correction for limb darkening, since the limb has the largest positive or negative Doppler shift and makes the largest relative contribution to the R-M signal \citep{Winn2010}.  Adopting the projected stellar rotation velocity $v\sin i\approx 14.4$ \kms{} (see Sec. \ref{sec:rotation}), we estimate $\Delta v_\mathrm{RM}$ to be $\approx 7$ m s$^{-1}$, which is a factor of 2-3 times smaller than the RV precision we achieved for individual frames. Visual inspection of the RV data during the transit did not reveal any significant feature indicative of an R-M effect (Fig. \ref{fig:rvs}).  We also investigated any variation of the mean line profile during the transit, termed the ``Doppler shadow" effect, by looking for changes or residuals in the cross-correlation function, but we found that, because of the lower-than-expected SNR of the original spectra, the noise was much larger than the expected signal of K2-100b's ``shadow", and we did not identify any transit signal.  Appendix \ref{sec:appendix} contains details of the RVs and cross-correlation analysis.

\subsection{10830\AA{} \hei\ Line}
\label{sec:hei}

\begin{figure}
	\includegraphics[width=\columnwidth]{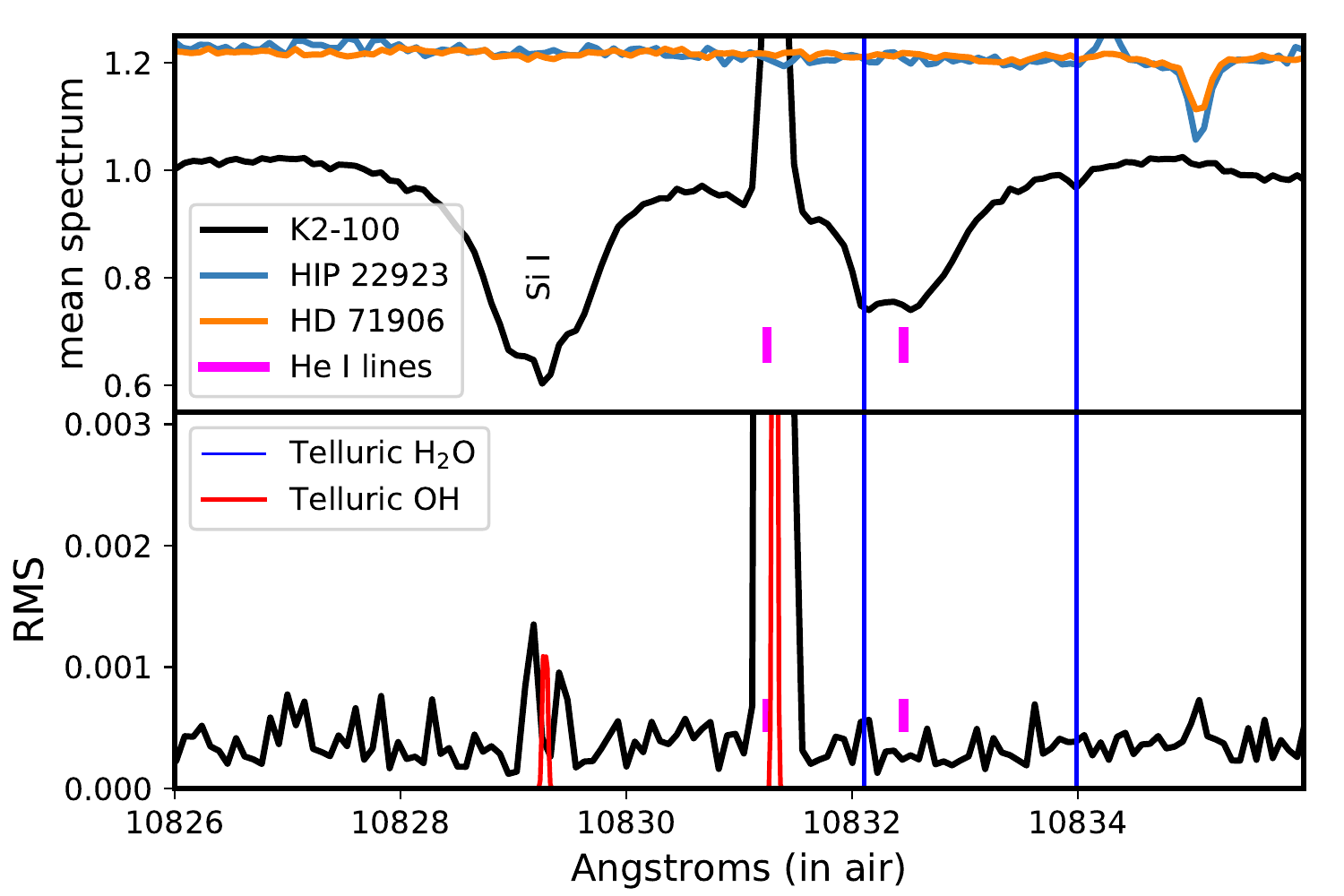}
    \caption{Top: Normalized average of spectra obtained of K2-100 and two A0-type comparison stars (offset for clarity).  Major stellar lines are labeled.  Bottom: Fractional standard deviation (RMS) of the spectra (black) compared to a model of sky (primarily OH) emission \citep{Noll2012,Jones2013} and the wavelengths of  telluric H$_2$O absorption lines identified by \citet{Breckinridge1973}.}  
    \label{fig:hei_rms}
\end{figure}

The average of all 15 spectra and the standard deviatory spectrum (RMS) in the vicinity of the triplet \hei\ lines are plotted in Fig. \ref{fig:hei_rms}.  
The theoretical airglow emission spectrum generated by {\tt SkyCalc} \citep{Noll2012,Jones2013} is also plotted in the RMS panel. The airglow spectrum is generated for an arbitrary sky condition, but we use this airglow spectrum only to identify the positions of telluric lines. 
There is a nearby strong stellar line of Si\,I, as well as telluric lines of OH and H$_2$O  in emission and absorption, respectively.  A strong OH line interferes with the weakest of the three \hei{} triplet lines, and a weak H$_2$O line contaminates the stronger two He I lines, producing the double line-like shape.

\begin{figure}
	\includegraphics[width=\columnwidth]{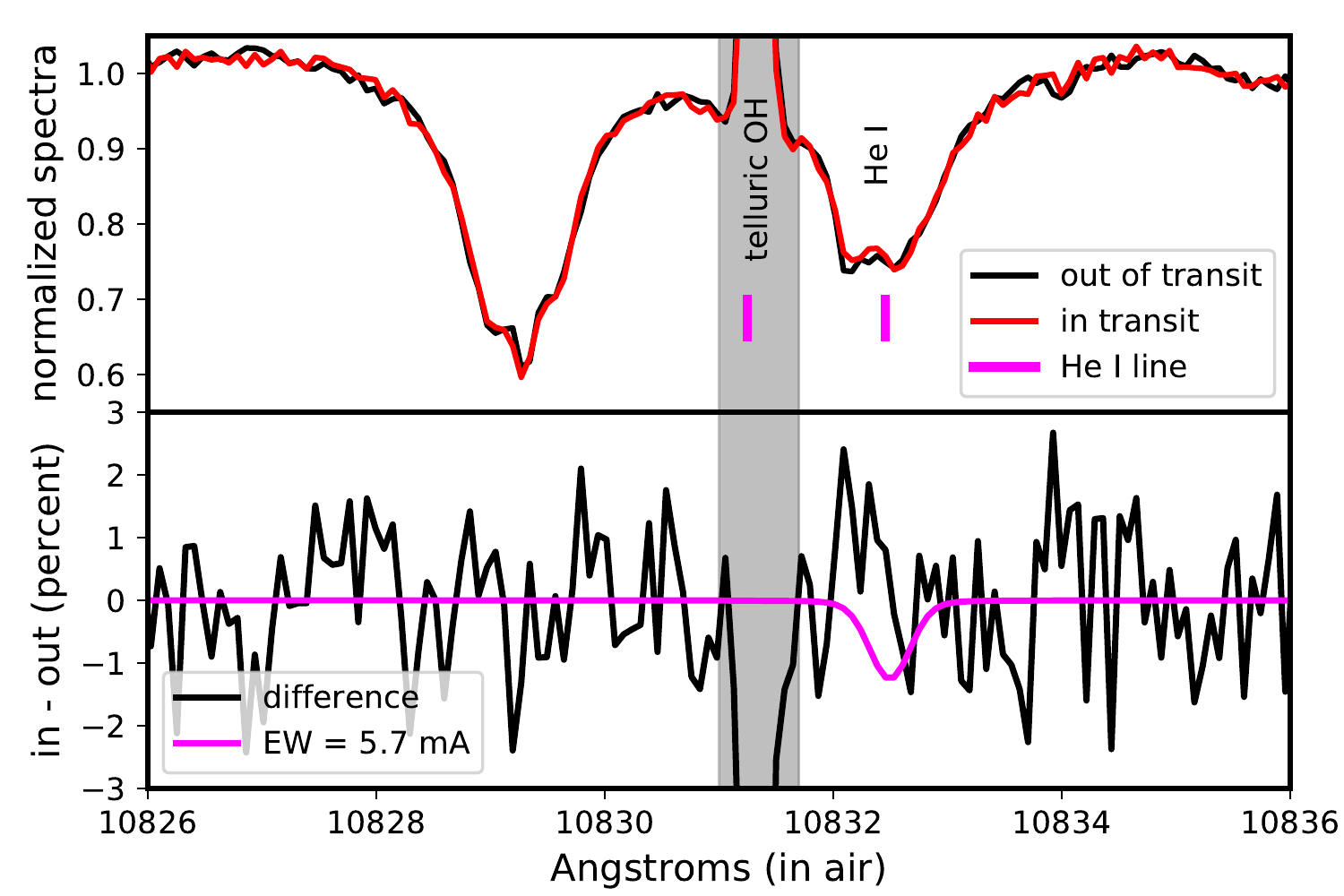}
    \caption{Top: Spectrum of K2-100 in the vicinity of the He I line during and outside of the transit of ``b".  The grey zone contains a strong telluric OH line and is excluded from analysis.  Bottom: The difference spectrum, compared to a model spectrum of a line with a total EW of 5.7 m\AA, the 99\% confidence detection limit.}  
    \label{fig:hei_line}
\end{figure}

Spectra of K2-100 were co-added during the transit (six spectra) and outside the transit (nine spectra).  In-transit spectra were normalized by the out-of-transit average, shifted to remove the planet's expected Doppler shift ($\pm 2\pi R_* \sqrt{1-b^2}/P = \pm 24$ \kms{} over the transit), and multiplied again by the out-of-transit average.  The two combined spectra are plotted in the top panel of Fig. \ref{fig:hei_line}.  The difference spectrum is plotted in the bottom panel of Fig. \ref{fig:hei_line}.  No transit-associated absorption is apparent.  To place a limit on any possible absorption we modeled the profile of all three He I lines for a range of total equivalent EW, computed a $\chi^2$ with a respect to the difference spectrum, and identified the EW corresponding to $p < 0.01$ with two degrees of freedom, i.e.  $\Delta \chi^2 < 9.21$.  We assumed Voigt profiles (see below) and thermal broadening of a 2000-20000K gas and adopted the standard deviation (1.1\%) of the difference spectrum in a 10\AA{} region of the neighborhood of the line, excepting the OH line, as the noise.   We find that the EW must be $<5.7$m\AA{} at 99\% ($p=0.01$) confidence, depending slightly on gas temperature, and plot the corresponding model as the red line in the bottom panel of Fig. \ref{fig:hei_line}.  A (variable) telluric H$_2$O line at 10832.109\AA{} could contributes systematic error of up several m\AA{} and increase this upper limit.   We also examined the Paschen $\alpha$ line of H I at 1.282 $\mu$m, an important indicator of accretion around young stellar objects \citep[e.g.,][]{Yasui2019} in the same fashion, but we saw no significant difference between the in-transit and out-of-transit spectra.

\section{The UV irradiation of K2-100b}
\label{sec:uv}

In Sec. \ref{sec:windmodel}, we translate our non-detection of transit-associated 10830\AA{} absorption into a limit on the amount of triplet-state \hei{} in a planetary wind and thence to a constrain on atmospheric mass loss rate.  The metastable triplet (2$^3$S) state of the \hei\ transition is primarily populated by recombination of He ionized by extreme ultraviolet (EUV) photons with energies $>26.4$ eV ($\lambda < 504$\AA) and primarily depopulated by ionizing near ultraviolet (NUV) photons with energies $>4.8$ eV \AA{} \citep[$\lambda < 2583$,][]{Oklopvcic2018}.  For this reason, interpretation of our observations requires knowledge of stellar emission and irradiation of its planet in the different UV bands.

\subsubsection{EUV}

The EUV is heavily extincted by the interstellar medium (ISM) and stars at the distance of the Praesepe cluster have not been detected at those wavelengths.  Moreover, the photochemistry of any planetary wind will be sensitive to the spectrum of the EUV emission, not just the overall flux.  Since K2-100 is a young, rapidly rotating star with elevated spot coverage and hence magnetic activity, we reconstructed its EUV spectrum as a scaled spectrum of the \emph{active} Sun as the only analog for which we have a detailed EUV spectrum.  We used the spectrum obtained on 28 February 2014 by the combination of the MEGS-A and MEGS-B channels of the Solar Dynamics Observatory (SDO)'s Extreme Ultraviolet Variability Experiment \citep[EVE,][]{Woods2012}.  This date was near the maximum of solar Cycle 24 and had the highest activity prior to the failure of the MEGS-A channel.  The maximum of Cycle 24 was subdued compared to previous cycles but was the only one observed with both channels of EVE, since SDO was only launched in 2010.  Since there are no direct EUV measurements for young, solar-type analogs of K2-100, we considered several observable proxies of the flux using nearby calibrator stars, or nearby stars that themselves can serve as analogs.   We considered four proxies: stellar disk-averaged emission in X-rays, line emission in the far ultraviolet (FUV), emission in the H and K lines of Ca II, and the stellar absorption line of \hei{} itself at 10830\AA.
  
{\it EUV from X-rays:}  \citet{Sanz-Forcada2011} used coronal models to construct a relationship between luminosity in X-rays (0.12-2.5 keV) as measured by the \emph{Roentge X-ray Satellite} (\emph{ROSAT}) and the EUV (100-920 \AA) luminosity.  \citet{Randich1995} report a detection of K2-100 with \emph{ROSAT} as source KW 100.  The estimated PSPC count-rate is $1.5 \pm 0.5 \times 10^{-3}$ sec$^{-1}$ which for a typical coronal temperature of 0.12 keV corresponds to a flux of $9.5 \pm 3.2 \times 10^{-15}$ ergs sec$^{-1}$ cm$^{-2}$ and, for a \gaia-based distance of 189~pc, $4.1 \pm 1.4 \times 10^{28}$ ergs sec$^{-1}$.  The relation of \citet{Sanz-Forcada2011} yields an EUV luminosity of $3.5 \pm 1.2 \times 10^{28}$ ergs sec$^{-1}$.  This may be a lower limit, since \citet{France2016} found that the \citet{Sanz-Forcada2011} coronal models consistently under-predict the contribution of chromospheric and transition regions to the EUV.  A second, indirect method of predicting EUV emission from X-ray emission uses the \lyalph{} emission at 1216 \AA{} as an intermediary.  The \lyalph{} line of K2-100 cannot be directly observed due to the ISM H~I over the intervening 189 pc.   Instead, we translated the bolometric-normalized X-ray luminosity of $\log R_X = -5.25 \pm 0.15$ into a bolometric-normalized Ly\,$\alpha$ luminosity using the relationship $\log R_{X} = 1.79 + (1.52 \pm 0.11 )\log R_{{\rm Ly}\alpha}$ of \citet{Landsman1993}, and got $R_{\rm Ly} = -4.65 \pm 0.35$, and $L_{{\rm Ly}-\alpha} = 2.2 \pm 1.8 \times 10^{29}$ ergs sec$^{-1}$.  We then translated this into an EUV (100-500 \AA) luminosity using the scaling relations of \citet{Linsky2014}, first by calculating a stellar-radius scaled Ly-$\alpha$ flux at 1 AU of $49 \pm 41$ ergs sec$^{-1}$ cm$^{-2}$, then summing the contributions from the first four lines in Table 5 of \citet{Linsky2014} to get a mean scaling factor of 0.45.  We arrived at a EUV luminosity of $\approx 1.1 \pm 1.2 \times 10^{29}$ ergs sec$^{-1}$.  This is nominally higher than the \citet{Sanz-Forcada2011}-based estimate and over a narrower wavelength range, but the uncertainty is very large.     

\emph{EUV from Ca II HK:} EUV emission has been related to emission in the H and K lines of singly-ionized calcium from active regions on the Sun \citep{Neupert1998}.  We calibrated this approximately linear relationship using the observed value for Ca II HK bolometric-normalized emission at the maximum of Cycle 24: $R'_{HK} = -4.93$ \citep{Egeland2017}.  \citet{Barragan2019} obtained $R'_{HK} = -4.45 \pm 0.01$ for K2-100, a factor of $3\times$ the solar value.  With the additional factor of K2-100's larger radius, this would lead to a 100-920\AA{} EUV luminosity of $5 \times 10^{28}$ ergs sec$^{-1}$, consistent with the estimate based on X-rays and the relation of \citet{Sanz-Forcada2011}.    

\begin{figure}
	\includegraphics[width=\columnwidth]{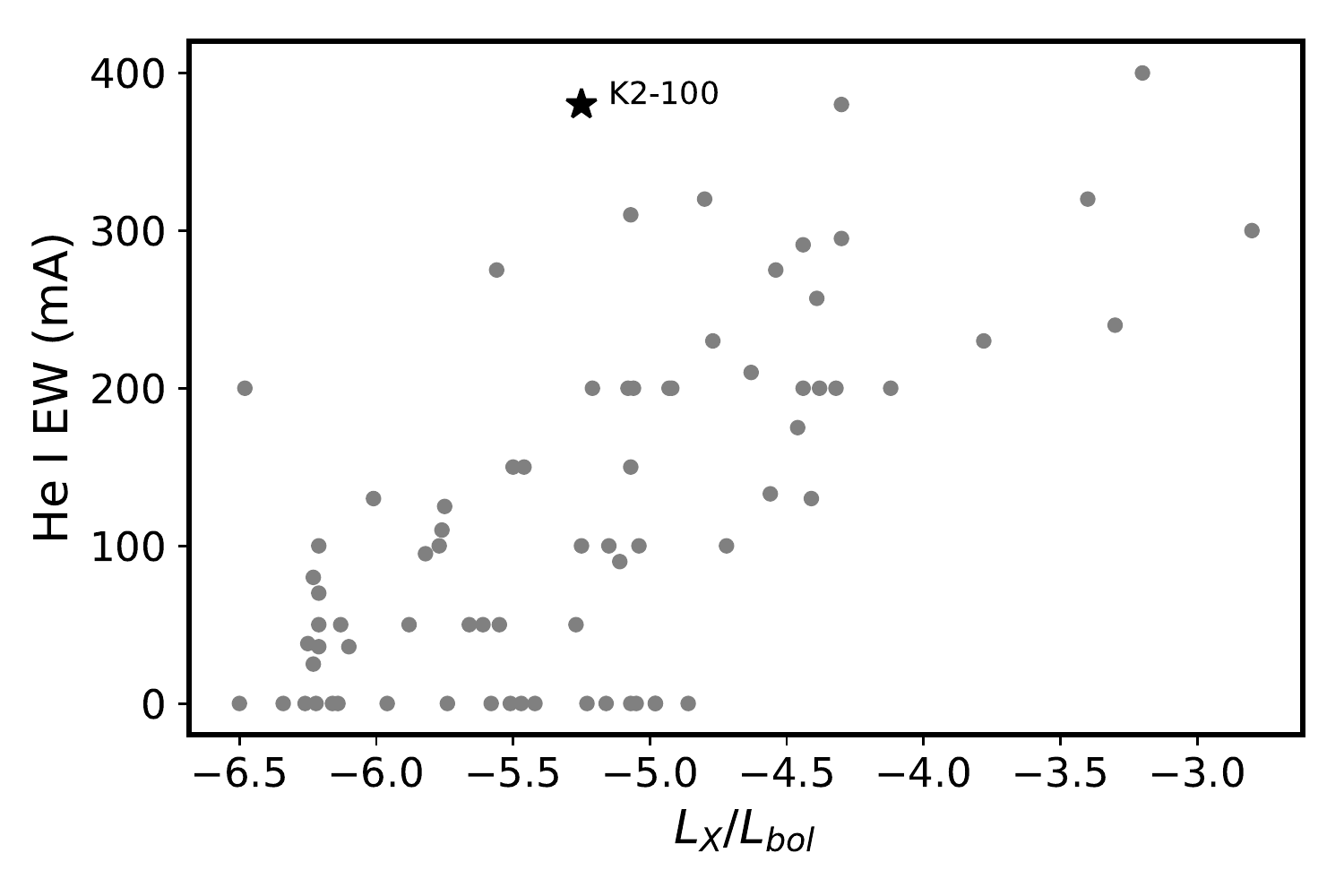}
    \caption{EW of the \hei{} line vs. normalized X-ray lumonsity of dwarf stars from \citet{Zarro1986} and \citet{Smith2016} compared to our observations of K2-100.} 
    \label{fig:smith}
\end{figure}

\emph{EUV from stellar \hei:} The He~I line of the stellar photosphere is also an indicator of stellar activity since the triplet helium in the lower state of the transition is produced by localized EUV emission from active regions on the star.  The advantage of this proxy is that the stellar line is measured during the planet's transit, obviating any effects of time variation in stellar magnetic activity.  \citet{Zarro1986} and \citet{Smith2016} found a correlation between X-ray emission and EW of the He I line.  \citet{Sanz-Forcada2008} found that the EW reached a maximum of about 400 m\AA{} and that the correlation disappeared for more active, X-ray luminous stars with $\log R_x > -4$, in part because \emph{emission} in the He I line reduced the EW.  We fit Voigt model line profiles to the \hei\ line complex plus a nearby Si I line at $\lambda = 10827.09$\AA{} (rest-frame in air) using the RV derived from the entire spectrum and fixing the Lorentz width but allowing the Gaussian broadening and EW of the lines to vary.  (See Sec. \ref{sec:windmodel} for more information on the line model).  The best-fit total EW of the three He I lines is 380 m\AA.  This value is higher than expected for the soft X-ray luminosity ($\log R_x = -5.25$) measured by \rosat{} (at a different epoch) compared to other dwarf stars with $B-V > 0.47$ (Fig. \ref{fig:smith}).  One possibility is that the X-ray luminosity during our IRD observations was much larger ($\log R_x \sim -4$), and that the EUV flux was concomitantly higher as well, although the scatter of the data in \citet{Smith2016} precludes a quantitative estimate.

\emph{EUV from Hyades analogs:} Finally, the EUV emission from K2-100 can also be estimated by identifying nearby analogs that are more readily studied.  There are few direct observations of EUV emission from normal stars because of the limited sensitivity of \emph{EUVE}, the only survey telescope dedicated to that energy regime \citep{SIrk1997}.  However, \citet{France2018} found a tight correlation between the EUV flux in 90-360 \AA{} and that in two FUV doublet lines: N V at 1238.82 and 1242.80\AA, and Si IV at 1393.75 and 1402.76 \AA.  Since the ISM is far less opaque in the FUV than the EUV, this permitted \citet{France2018} to estimate the EUV fluxes of many more nearby stars than previously ascertained by direct observation.  From the sample of \citet{France2018} we identified several stars (Table \ref{tab:analogs}) as potential analogs to K2-100 based on spectral type and Rossby number $Ro = P/\tau$, where $\tau$ is the global convective overturn timescale.  We use the $\tau$ vs. $B-V$ relationship developed by \citet{Mittag2018} combined with periods to estimate $Ro$.  The two most similar stars are probably HD 25825 and V993 Tau, members of the Hyades cluster which itself may have the same age and metallicity as Praesepe \citep{Brandt2015}.  Adopting the mean of the expected EUV emission of these two stars for K2-100, we predict a 90-360 \AA{} luminosity of $2.1 \pm 1.3 \times 10^{29}$ ergs~sec$^{-1}$.  This is higher than the proxy-based estimates, and over a comparatively narrower wavelength range, but we deem this approach to be a more direct and reliable determination than that using proxies.  We normalize the solar spectrum at $\lambda < 912$\AA{} such that the integrated flux in the 90-360\AA{} bandpass corresponded to that of the analog-based estimate.  (No stellar radius correction is applied since the analogs are expected to have radii similar to that of K2-100.)

\subsubsection{FUV and NUV}

As the basis for a FUV+NUV spectrum of K2-100 we took the annually-averaged spectrum of the Sun during the historically pronounced solar maximum of 1957 (Cycle 19), obtained from the LASP Interactive Solar Irradiance Datacenter (LISIRD), and scaled it with observed or inferred fluxes.  We estimated \lyalph{} emission of K2-100 based on the \citet{Landsman1993} relation with X-ray luminosity, i.e. $L_{{\rm Ly}-\alpha} = 2.2 \pm 1.8 \times 10^{29}$ (see above).  We scaled the emission in the FUV to that measured by the \galex{} satellite in its FUV channel \citep[$\lambda = 1350-1750$ \AA,][]{Boyce2007}.  We identified a source at the location of K2-100 in \galex{} image 3068361057647984640 (integration time 1560 sec) retrieved from the MAST Portal.  We measured the count rate in a 50-pixel circular aperture to be 0.080 sec$^{-1}$.  The background was determined as the scaled mean of a much larger aperture placed nearby and is 0.08 sec$^{-1}$.  We evaluated the error as the standard deviation of the counts to be 0.03 sec$^{-1}$ among nine 50-pixel apertures placed near the source apertures.  The flux density using the conversion of \citet{Morrissey2007} is $1.1 \pm 0.4 \times 10^{-16}$~ergs~sec$^{-1}$~cm$^{-2}$~\AA$^{-1}$ over a 255.5 \AA{} effective bandpass.  Combined with \gaia{} parallax this gives an FUV luminosity of $1.2 \pm 0.4 \times 10^{29}$~ergs~sec$^{-1}$.  We compared this with the value predicted by the the empirical relations with $B-V$ and $R'_{\rm HK}$ of \citet{Findeisen2011}, assuming a flat spectral intensity over the bandwidth.  This yields $L_{\rm FUV} = 1.7 \pm 0.5 \times 10^{29}$~ergs~sec$^{-1}$.   The scatter of stars about the best-fit FUV relation of \citet{Findeisen2011} is $\pm$30\%, thus the observed value and that predicted by the relation differ by only $1\sigma$.  K2-100 was not detected in a 96-sec image in the \galex{} near-ultraviolet (NUV, 1750-28000 \AA) channel, thus we used the \citet{Findeisen2011} relation to predict $1.0 \pm 0.2 \times 10^{32}$~ergs~sec$^{-1}$.   The predicted \galex\ magnitude $m_{\rm NUV} = 21.2$ is 0.7 magnitudes fainter than the limiting sensitivity for the 96-second integrations of the All-sky Imaging Survey \citep{Martin2005}, explaining the non-detection of the star.

Table \ref{tab:irradiance} lists the estimated luminosities of K2-100 and irradiance of K2-100b in the different wavelength regimes, assuming a circular orbit. 

\begin{table}
\begin{center}
\caption{Nearby Analogs of K2-100 from \citet{France2018} \label{tab:analogs}} 
\begin{tabular}{lllllll}
\multicolumn{1}{c}{Name} & \multicolumn{1}{c}{SpT} & \multicolumn{1}{c}{Age (Gyr)} & \multicolumn{1}{c}{Period} & \multicolumn{1}{c}{$\tau_{\rm conv}$} & \multicolumn{1}{c}{Ro} & \multicolumn{1}{c}{$F_{\rm EUV}^e$}\\
\hline
 & & \multicolumn{1}{c}{(Gyr)} & \multicolumn{1}{c}{(days)} & \multicolumn{1}{c}{(days)} &  & \multicolumn{1}{c}{}\\
\hline
$\tau$ Boo  & F7 & 1.6-2.3$^{f}$ & 3.31 & 15.0 & 0.22	& 7.77\\
$\zeta$ Dor & F9 & 0.6$^{g}$ & 3.43$^{a}$ & 27.6 & 0.12 & 2.46\\
HD 25825    & G0 & 0.8$^{c}$ & 7.47$^b$ & 27.8 & 0.27 & 9.42 \\
V993 Tau    & G0 & 0.8$^{c}$ & 5.87$^d$ & 20.2 & 0.29 & 7.49 \\
\hline
\end{tabular}
\end{center}
$^{\rm a}$from analysis of \emph{TESS} photometry.\\
$^{\rm b}$from \citet{Douglas2019} and analysis of ASAS-SN photometry.\\
$^{\rm c}$member of the Hyades cluster.\\
$^{\rm d}$from \citet{Radick1995}.\\
$^{\rm e}$ $10^4$ ergs~sec$^{-1}$~cm$^{-2}$ (90-360\AA).\\
$^{\rm f}$\citet{Mamajek2008}.\\
$^{\rm g}$\citet{Maldonado2012}.
\end{table}

\begin{table*}
\begin{center}
\caption{Irradiation of K2-100b \label{tab:irradiance}} 
\begin{tabular}{llrrl}
\multicolumn{1}{c}{regime} & \multicolumn{1}{c}{$\lambda\lambda$} &  \multicolumn{1}{c}{luminosity} & \multicolumn{1}{c}{flux at K2-100b} & \multicolumn{1}{c}{source}\\
\hline
 & \multicolumn{1}{c}{\AA} & \multicolumn{1}{c}{10$^{30}$ ergs~sec$^{-1}$} & \multicolumn{1}{c}{ergs~sec$^{-1}$~cm$^{-2}$} &\\
\hline
bolometric &  & $7390 \pm 190$ & $2.98 \pm 0.08 \times 10^9$ & \citet{Mann2017} + \gaia{} DR2\\
X-ray & 5-120 & $0.041 \pm 0.014$ & $1.7 \pm 0.6 \times 10^4$ & \emph{Rosat} PSPC\\
EUV & 90-360 & $0.21 \pm 0.13$ & $8.5 \pm 5.0 \times 10^5$ & \citet{France2018} analogs\\
Ly~$\alpha$ & 1180-1250 & $0.21 \pm 0.18$ & $8.5 \pm 7.0 \times 10^5$ & \citet{Landsman1993} relation\\
FUV & 1350-1750 & $0.12 \pm 0.4$ & $4.8 \pm 1.6 \times 10^4$ & \galex{} DR6/7\\
NUV & 1750-2800 & $100 \pm 20$ & $4.0 \pm 0.8 \times 10^7$ & \citet{Findeisen2011} relation\\
\hline
\end{tabular}
\end{center}
\end{table*}

\section{Planetary Wind Model}
\label{sec:windmodel}

To translate our upper limit on 10830\AA{} \hei{} line absorption associated with the transit of K2-100b into statements about mass loss, we modeled the escaping atmosphere of the planet and its photochemistry as a spherically symmetric, isothermal Parker wind \citep{Parker1958} with temperature $T_w$.  Our model largely recapitulates the model of \citet{Oklopovcic2018}, and we refer the reader to that work for the details, however, we calculate the wavelength-dependence optical depths and photoionization rates explicitly, without averages as approximations.  Temperature-dependent recombination coefficients for single and triplet He are from \citet{Osterbrock1989}, collision induced transition strengths and energies are from \citet{Bray2000}, and neutral collisional de-excitation coefficients are from \citet{Roberge1982}.  We assumed a solar-like composition for the wind of 91.3\% H and 8.7\% He by number.  Line profiles were calculated using oscillator strengths from \citet{Wiese2009}, assuming Voigt profiles, and integrating over the accessible wind out to the Roche radius, while accounting for the Doppler shift due to the projected wind velocity.  These intrinsic profiles do not include the effect of the finite resolution of the spectrograph ($\approx 4$ \kms), or the orbital motion of the planet during the transit, which will blur profiles by $2\pi R_* t/ (P \tau)$ or about 1.5 \kms{}, since these are small due to the expected thermal broadening ($>6$ \kms).

\begin{figure}
	\includegraphics[width=\columnwidth]{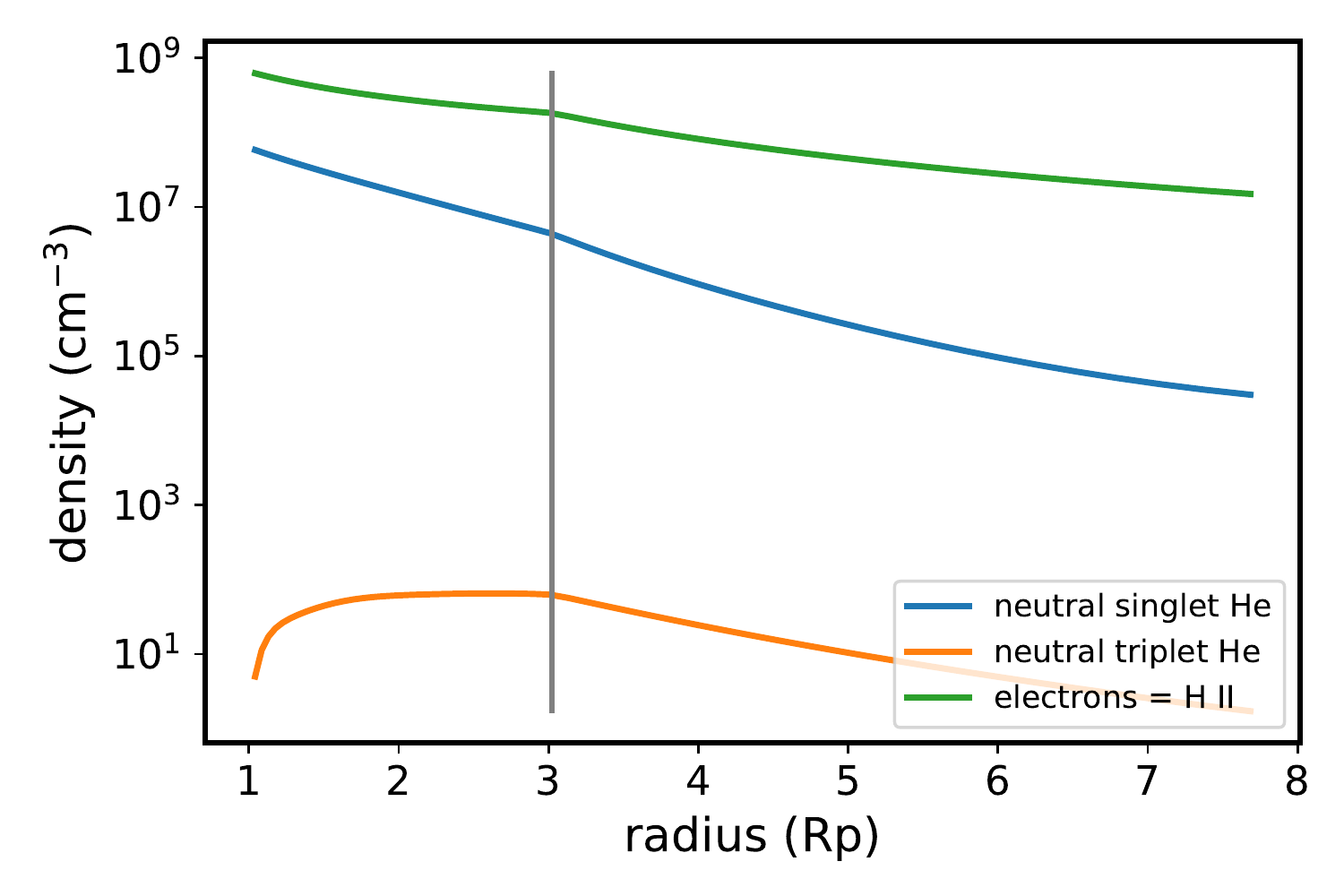}
    \caption{Density profiles of electrons (green line) and \hei{} in the singlet and triplet states in a model of an isothermal Parker wind from K2-100b.  H is essentially entirely ionized by stellar EUV photons and not shown.  The vertical grey line delineates the sonic point.  The conditions are $\dot{M} = 1M_{\oplus}$~Gyr$^{-1}$ and $T_w = $10000K.}  
    \label{fig:densities}
\end{figure}

Figure \ref{fig:densities} plots the densities of electrons (equivalent to the density of ionized H), and He I in the singlet and triplet electronic configurations for the case of $\dot{M} = 1$\mearth Gyr$^{-1}$ and $T_w = 10000$\,K.  Escaping neutral hydrogen from K2-100b is rapidly and completely ionized by EUV photons from K2-100.  We can understand the model prediction that H\,I is completely ionized (except very close to the planet) by performing a simple Str\"{o}mgren-sphere-like comparison which competes the rate of impinging photoionizing photons against the rate of escaping H\,I atoms:
\begin{equation}
\label{eqn:stromgren}
    I_{\rm EUV} \pi R_s^2 > \dot{M} X_H / \mu.
\end{equation}
We estimate the rate of impinging photoionizing photons (the left hand side) to be $3 \times 10^{35}$~sec$^{-1}$, while the escape rate of H I atoms for a mass loss of 1 \mearth{} Gyr$^{-1}$ is $8 \times 10^{34}$~sec$^{-1}$.  Although this simple calculation (but not the model) ignores recombination, it shows that the EUV irradiation field at the orbit of K2-100b is capable of completely ionizing the flow of H.  (We verified that at substantially lower EUV irradiance, the model predicts significant H\,I in the wind.)  

\begin{figure}
	\includegraphics[width=\columnwidth]{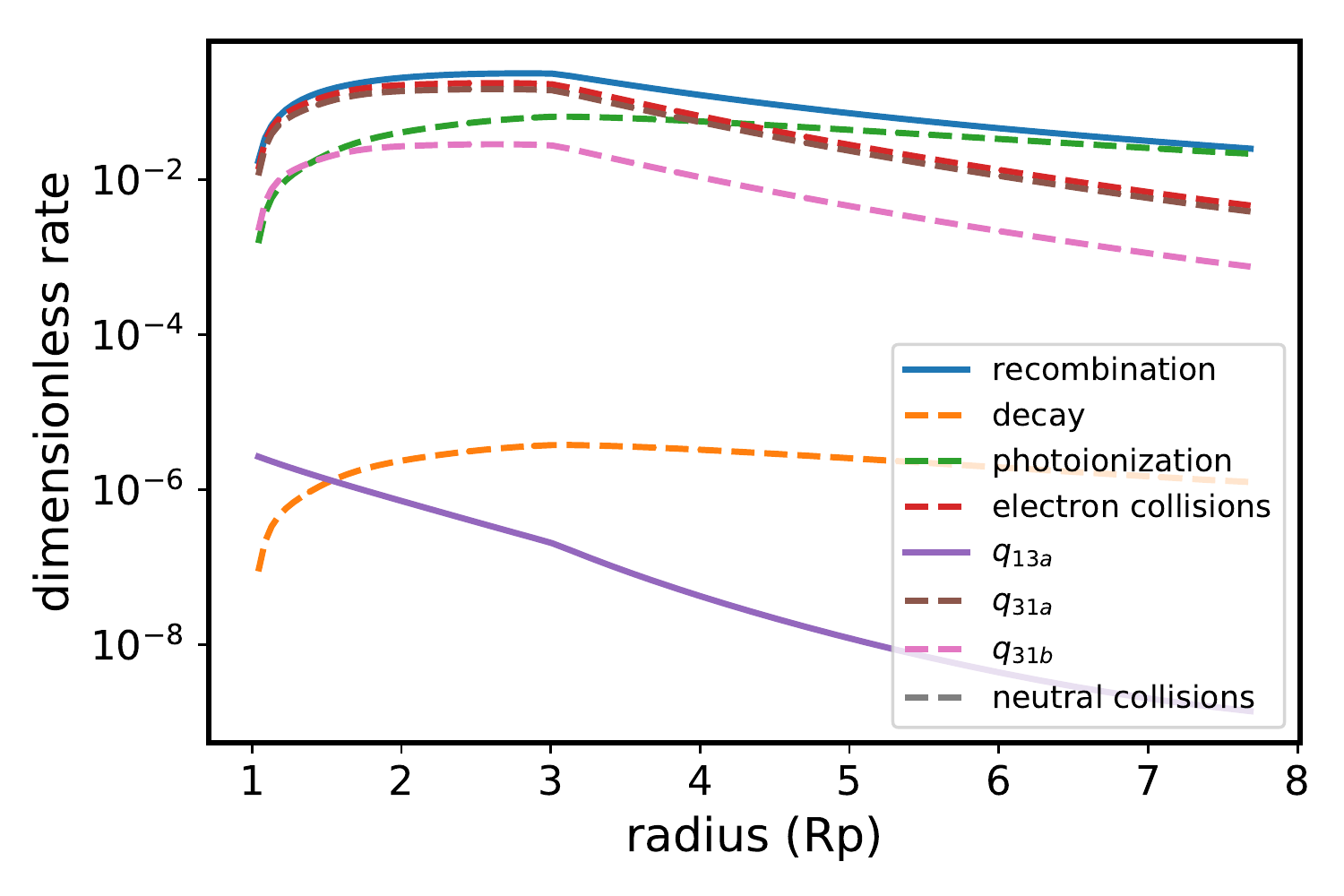}
    \caption{Rates of processes that populate (solid lines) or depopulate (dashed lines) the 2s$^{3}$S triplet state in the 10830\AA{} transition of \hei, normalized by the timescale $R_s/c_s$, where $R_s$ and $c_s$ are the sonic radius and speed, respectively.  The conditions are mass loss rate $\dot{M} = 1M_{\oplus}$~Gyr$^{-1}$ and wind temperature $T_w = $10000K.}  
    \label{fig:rates}
\end{figure}

Figure \ref{fig:rates} plots the rates of processes that populate (solid lines) or deplete (dashed lines) the 2s$^{3}$ ground state of the 10830 \AA{} line, normalized by the wind timescale $R_s/c_s$, where $R_s$ is the sonic radius and $c_s$ is the isothermal sound speed.  ``Electron collisions" is the \emph{net} depopulation of the 2s$^{3}$S level by electron collisions.  The 2s$^{3}$S  state is almost entirely populated by recombination of ionized He (blue solid line) and depopulated by a combination of electron collisional de-excitation (dashed brown line), dominant in the inner part of the wind, and photoionization (dashed green line), dominant in the outer part.  \citet{Oklopvcic2018} find that photoionization is pervasively dominant, but used a mass loss rate and hence densities that are five times lower than that assumed here.    

\begin{figure}
	\includegraphics[width=\columnwidth]{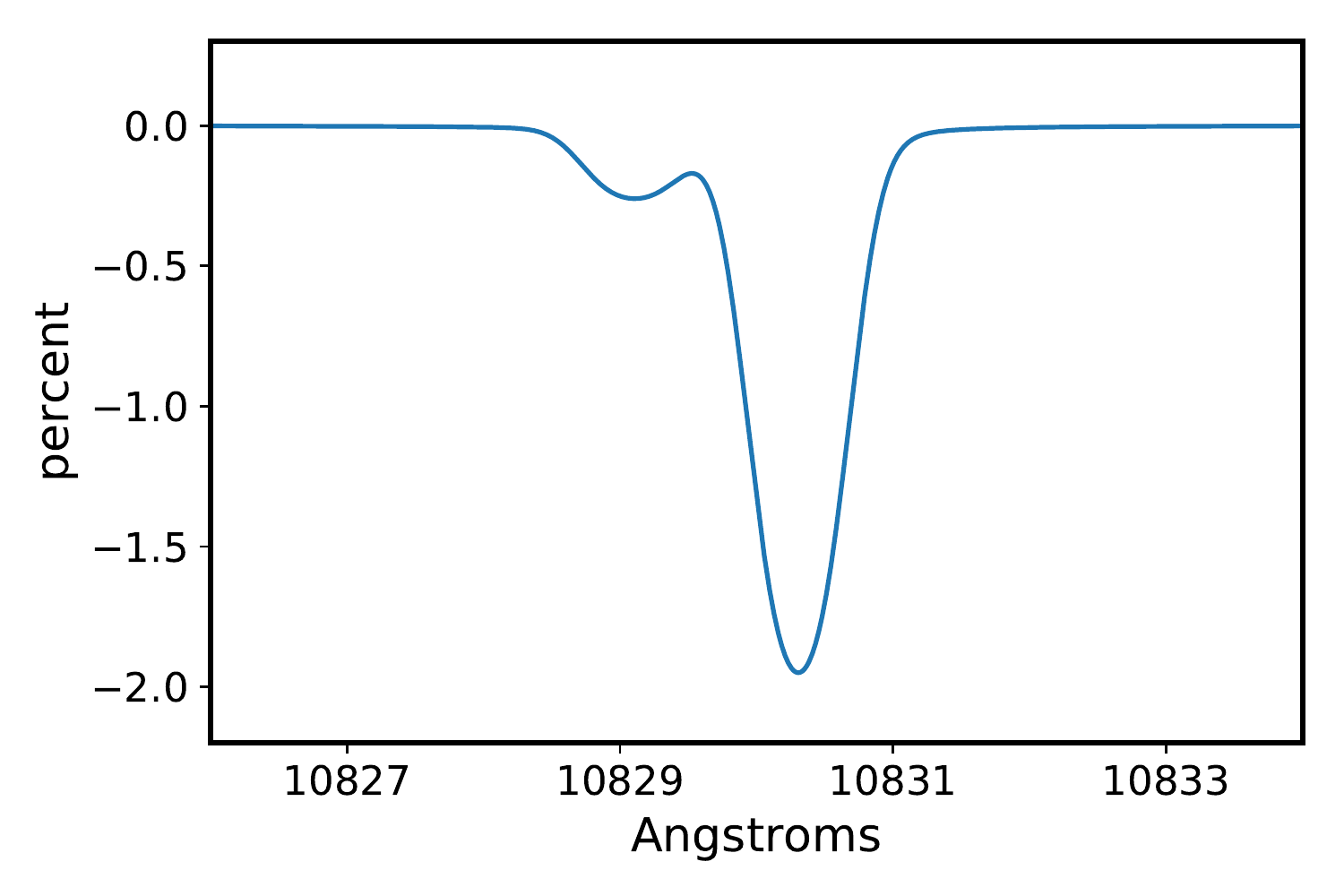}
    \caption{Absorption spectrum of a planetary wind in the \hei{} 10830\AA{} line, expressed as 
\% of the out-of-transit spectrum.  The conditions are $\dot{M} = 1M_{\oplus}$~Gyr$^{-1}$ and $T_w = $10000K.  Wavelengths are as observed in air and no Doppler shift is imposed.  The EW is 18 m\AA{}, which would have been readily detected in our IRD observations of the transit of K2-100b.}  
    \label{fig:profile}
\end{figure}

Figure \ref{fig:profile} shows the calculated profile of the triplet \hei{} line complex produced by a planetary wind with $\dot{M} = 1$\mearth Gyr$^{-1}$ and $T_w = 10000$\,K.  The total EW is 18 m\AA{} and the two strongest lines of the the three are blended by Doppler broadening ($\approx$15 \kms{} or about 0.5\AA).  

\begin{figure}
	\includegraphics[width=\columnwidth]{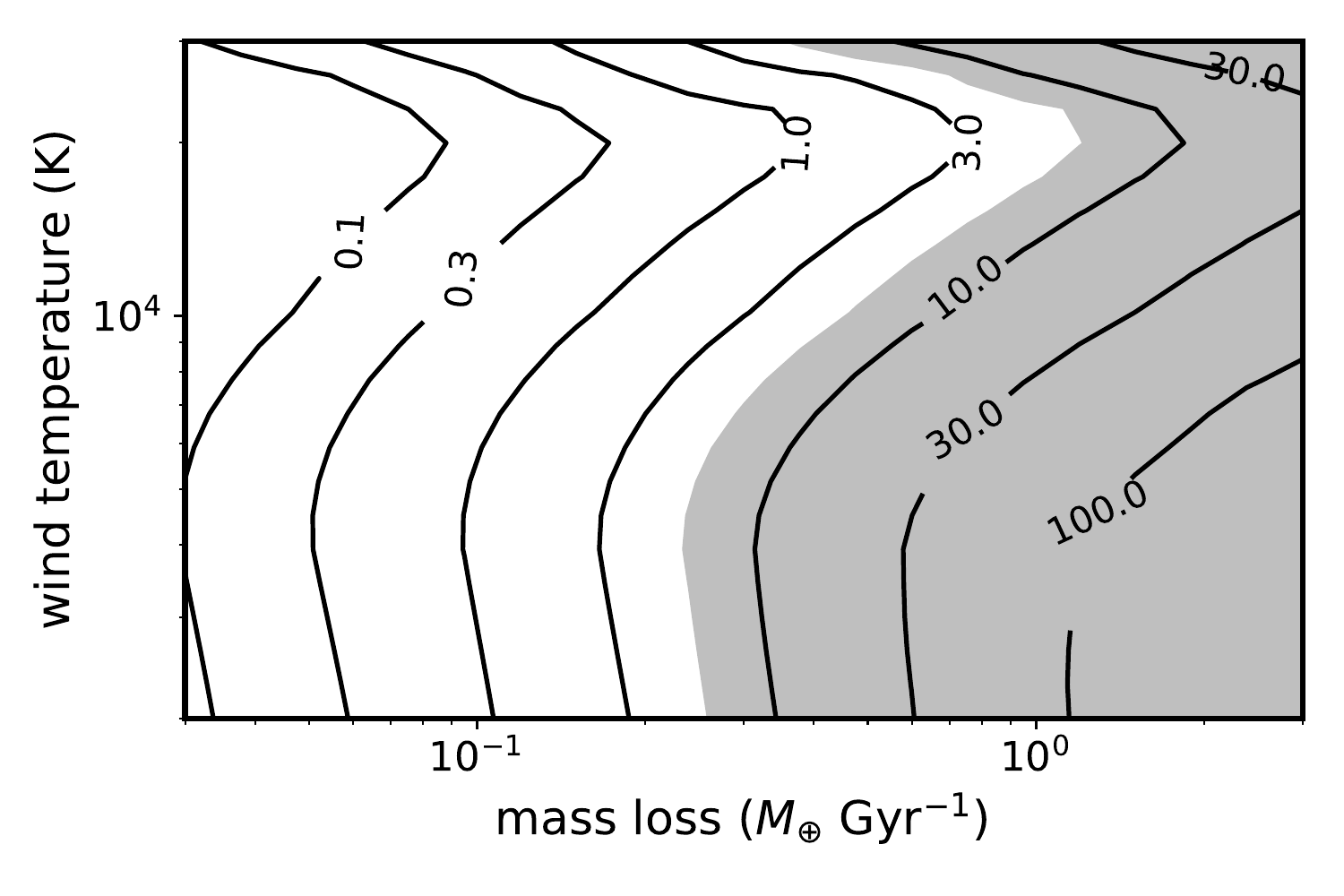}
    \caption{Predicted combined EW (in m\AA) of the 10830 \AA{} triplet \hei{} line vs. mass loss in Earths per Gyr ($\approx 2 \times 10^9$ kg~sec$^{-1}$) and wind temperature.  The grey area is excluded at 99\% confidence by the upper limit of 5.7 m\AA.}  
    \label{fig:hei}
\end{figure}

These model calculations suggest that for a solar-like composition, an upper limit of total EW $<5.8$ m\AA{} rules out mass loss rates $>$1 \mearth{} Gyr$^{-1}$ for all wind temperatures \twind{} considered here, and loss rates $>$0.3 \mearth{} Gyr$^{-1}$ for \twind$<10000$K (Figure \ref{fig:hei}).  \hei{} line observations are maximally sensitive to winds with \twind=4000-5000K, and least sensitive to those at 20000K.  Wind temperature affects the sound speed, the sonic point, recombination rates, collisional de-excitation and excitation rates.  For a given $\dot{M}$, increasing \twind{} decreases density and hence the EW of the line.  The recombination rate coefficients of \citet{Osterbrock2006} also decrease with temperature and the density is lower so the actual rate is lower still.  Photoionization, the primary mechanism that depopulates the triplet state, is independent of \twind.  The ``a" pathway of collisional de-excitation ($q_{31}a$) peaks at 18000K, and this may explain some of the high-temperature structure in Fig. \ref{fig:hei}, but we caution that both the low- and high-\twind{} calculations depend on extrapolated values for the recombination rates of ionized He to singlet and triplet \hei{} and may not be reliable.   

\section{Interpretation and Discussion}
\label{sec:interpretation}

The EUV and Ly $\alpha$ emission of K2-100 have not been measured and presumably will never be directly measured due to the neutral H~I along the intervening 186 pc to this star.  Both the X-ray and FUV emission used to estimate those fluxes were only marginally ($3\sigma$) detected and \emph{formal} uncertainties in estimates of the EUV flux are of order unity.  It is therefore prudent to consider the sensitivity of the modeling results to changes in the EUV emission at different wavelengths.  We calculated the EW of the 10830 \AA{} \hei{} line for a mass loss rate of 1 \mearth{} Gyr$^{-1}$ and wind temperature of 10000K, and repeated the calculation while doubling the UV irradiance in 10 \AA-wide windows over the entire spectral region, recording the fractional change in the calculated EW.  We did \emph{not} vary the mass loss rate, which of course, would also be expected to change with varying EUV flux.  Rather, this numerical experiment examines our ability to robustly translate an observed EW (or an upper limit) into a given mass-loss rate and indicates at wavelengths it is most important to establish the stellar emission. 

\begin{figure}
	\includegraphics[width=\columnwidth]{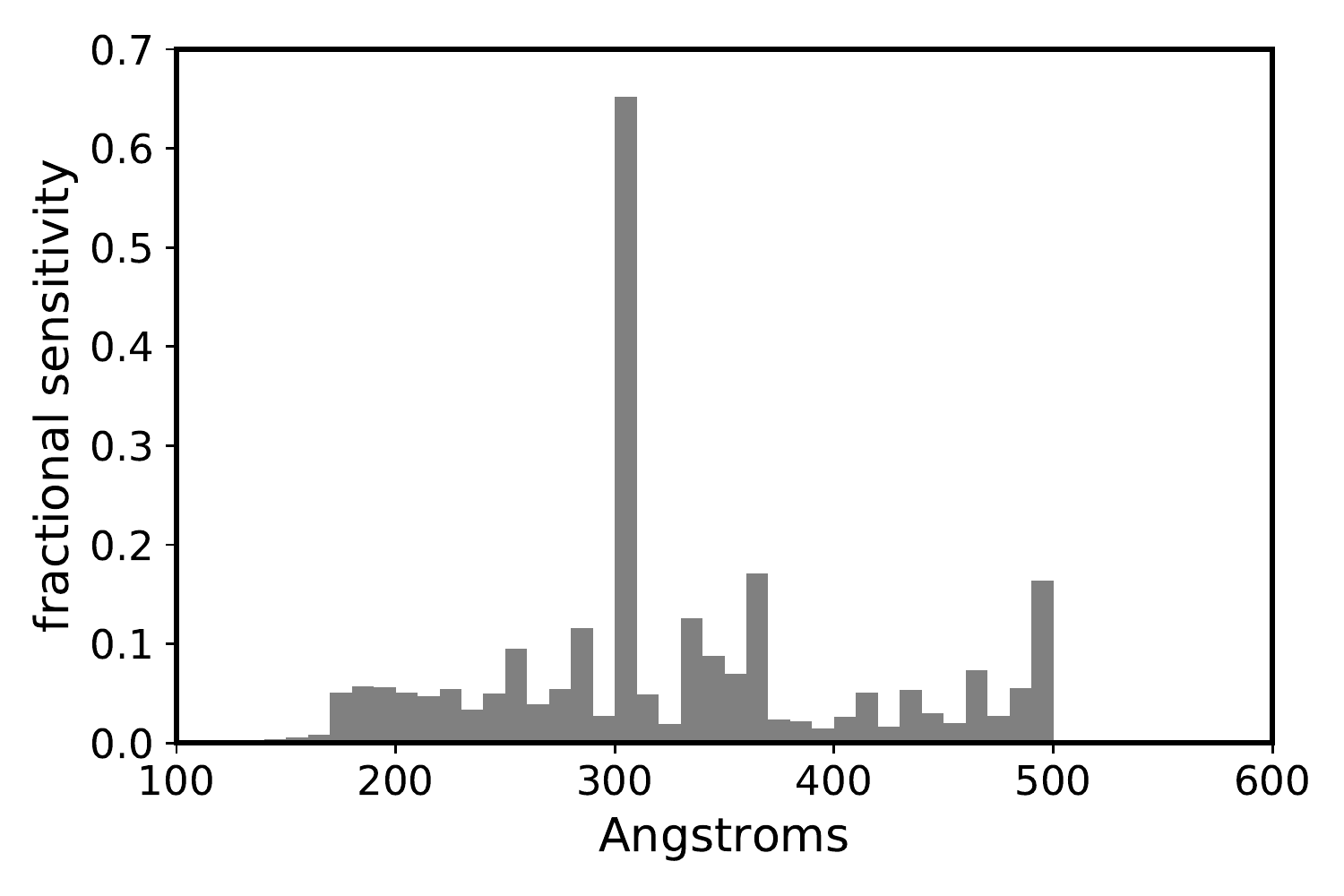}
    \caption{Fractional change in the EW of the \hei{} line at 10830 \AA{} in response to a doubling of the EUV emission in 10\AA{} wavelength bins.  Conditions are a mass loss rate of 1 \mearth{} Gyr$^{-1}$ and wind temperature of 10000K.}  
    \label{fig:euv_sensitivity}
\end{figure}

Figure \ref{fig:euv_sensitivity} shows the results for a restricted range in the EUV where the sensitivity was the non-negligible.  Outside this range, the sensitivity was $<1$\%, except for a range in the NUV where increasing irradiance produced a slightly lower EW due to elevated photoionization.  In essence, this is a convolution of the EUV spectrum of the star (in this case, a scaled version of the Solar spectrum) and the photoionization cross-section of \hei{}.  This shows that it is most important to accurately measure the EUV emission in the 15-500\AA{} range, and in particular at the 305\AA{} He II resonant line for which there is already strong emission from Sun-like stars \citep{Golding2017}.  Our planetary wind models show that the 2s$^{3}$S state of \hei\ is primarily produced by recombination of ionized helium, and the range of non-negligible sensitivity corresponds to the range over which He I ionization occurs ($\lambda < 504$\AA).  This result should serve as a guide for where future efforts to estimate the EUV emission from stars for the purpose of He I line observation should concentrate. 
\citet{Oklopvcic2019} discusses how the \hei{} line strength is governed by the balance between EUV radiation, which produces 2s$^{3}$S \hei\ via ionization of the ground state and subsequent recombination, and mid-UV radiation, which ionizes tripleet \hei.  For a fixed mass loss rate, wind temperature, and separation from the host star, there is a spectral type for which the triplet-state \hei\ signal is maximum; hotter stars have more mid-UV emission, and cooler stars do not ionize enough ground-state \hei.  For $\dot{M}$ of 0.2 \mearth~Gyr$^{-1}$ and a wind temperature of 7300K this optimal spectral type is K-type, consistent with the observation that all four successful detections of \hei\ are for planets orbiting K-type stars, and there are only non-detections for planets transiting A-, G-, and M-type stars.  Two planets which were not detected, WASP-12b of \citet{Kreidberg2018} and HD 209485b of \citet{Nortmann2018}, orbit G0 stars like K2-100, and are at separations which are well within a factor of two of K2-100b.  However, those stars are older, less magnetically active, and presumably would have lower EUV luminosities.  The upper limits for WASP-12b and HD209458b, converted into EW, are 24 m\AA{} (90\%) and 8 m\AA{}, respectively (both $2\sigma)$, comparable to our formal 2.5$\sigma$ limit of 5.7 m\AA.

We can use the UV irradiances calculated and reported in Table \ref{tab:irradiance} to estimate the energy-limited escape rate $\dot{M}_{\rm EL}$ of the envelope of K2-100b, an upper limit on the escape rate in the absence of other energy sources and ignoring processes such as erosion by the stellar wind.  The relation \citep{Watson1981,Erkaev2007} is:
\begin{equation}
\dot{M}_{\rm EL} = \frac{\eta \pi F_{\rm XUV}R_{\rm XUV}^3}{KGM_p}
\end{equation}
where $\eta$ is an efficiency factor that accounts for energy loss by radiative and conductive cooling, ionization, as well as the remaining heat content of the escaping gas, $F_{\rm XUV}$ is the combination of X-ray and UV irradiances, $R_{\rm XUV} > R_p$ is the effective radius at the level fo the atmosphere which the XUV radiation is absorbed and from which escape occurs, $G$ is the gravitational constant, and $K$ is a correction factor $\approx 1$ that accounts for the fact that the escaping gas only needs to reach the Roche radius \citep{Erkaev2007}.  Adopting $\eta = 0.1$ \citep{Shematovich2014} and $R_{\rm XUV} = R_p$, and the mass estimate of \citet{Barragan2019}, we obtain an estimate for energy-limited escape rate of 1.1 \mearth{}\,Gyr$^{-1}$ driven by the combination of X-ray, EUV, and Ly $\alpha$ irradiation.  This scenario is ruled out by our observations.  X-rays, EUV, and Lyman $\alpha$ may drive drive atmospheric escape with different efficiencies depending on proximity to the host star \citep{Owen2012}, but X-rays  are a small fraction of the total energy budget.   Asymmetry of the escaping atmosphere may also affect our ability to observe triplet \hei{} and thus limit mass loss.  Otherwise, the efficiency of escape could be low or the irradiance significantly lower than estimated.  Our analysis of K2-100b's orbit based on transit observations limit the eccentricity to 0.15 (90\% confidence) or 0.28 (99\% confidence) and do not support an origin for this object as a more massive planet on a scattered orbit.   Perhaps photoevaporative escape, expected to be important during the first few hundred Myr when the star was rapidly rotating and magnetically active \citep{Owen2019}, has largely halted, leaving the planet stranding in the desert.  Alternatively, K2-100b's atmosphere is escaping at a lower rate but over a longer (Gyr) interval, consistent with the scenario of core accretion-powered escape \citep{Ginzburg2016}.

\section*{Acknowledgements}

We thank Evgenya Shkolnik for a helpful discussion on stellar UV emission.  EG acknowledges support as a visiting professor from the German Science Foundation through the DFG Research Unit FOR2544 ``Blue Planets around Red Stars". A.W.M. and D.O. were supported by the \ktwo\ Guest Observer Program (NASA grant 80NSSC19K0097) to the University of North Carolina at Chapel Hill. This work is supported by JSPS KAKENHI Grant Numbers 16K17660, 19K14783, 18H05442, 15H02063, and 22000005, and by the Astrobiology Center Program of the National Institutes of Natural Sciences (NINS) (Grant Number AB311017).  This research has made use of the SVO Filter Profile Service (http://svo2.cab.inta-csic.es/theory/fps/) supported from the Spanish MINECO through grant AYA2017-84089, the  LASP interactive Solar Irradiance Datacenter, NASA's Astrophysics Data System Bibliographic Services, Centre de Donn\'{e}es astronomiques de Strasbourg, NIST's atomic line database, {\tt Astropy} \citep{Astropy2013}, and {\tt Scipy} \citep{Scipy2019}.





\begin{table*} 
\caption{Parameters of K2-100b\label{tab:param}} 
\begin{tabular}{lcc} 
\hline 
\hline 
Parameter & \multicolumn{2}{c}{Value} \\ 
\hline
 & e float & e fixed at 0 \\
\hline 
\multicolumn{3}{c}{Measured Parameters} \\ 
$T_0$(BJD-2454833) &  $2311.06731^{+0.00037}_{-0.00053}$ & $2311.06738 \pm 0.00015$ \\ 
$P$ (days) & $1.67390323 \pm 2.5\times10^{-7}$ & $1.67390323 \pm 2.4\times10^{-7}$ \\ 
$R_P/R_{\star}$ & $0.02883^{+0.00023}_{-0.00026}$ & $0.02876^{+0.00024}_{-0.00025}$ \\ 
$b$ & $0.767^{+0.016}_{-0.02}$ & $0.76^{+0.019}_{-0.022}$\\ 
$q_{1,{\it Kepler}}$ & $0.284^{+0.028}_{-0.025}$ & $0.285^{+0.028}_{-0.026}$ \\ 
$q_{2,{\it Kepler}}$ & $0.366^{+0.08}_{-0.092}$ &$0.362^{+0.083}_{-0.096}$ \\ 
$q_{1,SDSS i'}$ & $0.101^{+0.123}_{-0.073}$ &  $0.108^{+0.126}_{-0.077}$ \\ 
$q_{2,SDSS i'}$ & $0.31^{+0.11}_{-0.12}$ & $0.31^{+0.11}_{-0.12}$\\ 
$q_{1,SDSS r'}$ & $0.38^{+0.21}_{-0.2}$ & $0.37^{+0.19}_{-0.18}$ \\ 
$q_{2,SDSS r'}$ & $0.352^{+0.091}_{-0.1}$ & $0.352^{+0.09}_{-0.1}$ \\ 
$q_{1,SDSS z'}$ & $0.29^{+0.17}_{-0.16}$ & $0.29^{+0.18}_{-0.16}$\\ 
$q_{2,SDSS z'}$ & $0.32^{+0.11}_{-0.13}$  & $0.32^{+0.11}_{-0.13}$ \\ 
$\sqrt{e}\sin\omega$ & $0.085^{+0.089}_{-0.088}$& 0[fixed]\\ 
$\sqrt{e}\cos\omega$ & $-0.06 \pm 0.26$ & 0[fixed] \\ 
\hline 
\multicolumn{3}{c}{Derived Parameters} \\ 
$a/R_{\star}$ & $5.34^{+0.12}_{-0.11}$ & $5.53^{+0.2}_{-0.18}$ \\ 
$i$ ($^{\circ}$) & $81.51^{+0.32}_{-0.33}$ & $82.1^{+0.51}_{-0.47}$\\ 
$\delta$ (\%) & $0.0831^{+0.0013}_{-0.0015}$ & $0.0827 \pm 0.0014$ \\ 
$g_{1,{\it Kepler}}$ & $0.389^{+0.079}_{-0.092}$ & $0.385^{+0.081}_{-0.097}$\\ 
$g_{2,{\it Kepler}}$ & $0.142^{+0.105}_{-0.085}$ &$0.147^{+0.109}_{-0.09}$\\ 
$g_{1,SDSS i'}$ & $0.175^{+0.117}_{-0.091}$ &$0.183^{+0.126}_{-0.098}$\\ 
$g_{2,SDSS i'}$ & $0.108^{+0.129}_{-0.074}$& $0.111^{+0.127}_{-0.075}$ \\ 
$g_{1,SDSS r'}$ &  $0.41 \pm 0.17$ & $0.4^{+0.17}_{-0.16}$ \\ 
$g_{2,SDSS r'}$ &  $0.17^{+0.15}_{-0.11}$ &  $0.17^{+0.15}_{-0.11}$ \\ 
$g_{1,SDSS z'}$ & $0.32^{+0.17}_{-0.16}$ &  $0.32^{+0.18}_{-0.16}$\\ 
$g_{2,SDSS z'}$ & $0.18^{+0.15}_{-0.11}$ & $0.18^{+0.16}_{-0.11}$ \\ 
$e$ & $0.054^{+0.088}_{-0.037}$ & 0 \\ 
$\omega$ ($^{\circ}$) & $136.2^{+44.7}_{-100.0}$  & 0\\ 
\hline 
\end{tabular} 
\end{table*} 

\appendix
\section{Analyses of the Line Profile}
\label{sec:appendix}
The mean profile of spectral lines  contains information on the rotation, atmospheric turbulence, and magnetic activity of the star \citep{Gray2005}.  During a transit, the line profile variation against time also allows us to investigate the orbital trajectory of the transiting planet with respect to the stellar spin, and thus measure the stellar obliquity \citep[e.g.,][]{CollierCameron2010}. To measure the projected rotation velocity and stellar obliquity for K2-100, we analysed the line profile (variation) for the IRD spectra taken on UT 2019 March 24.   Following \citet{2020ApJ...890L..27H} we exploited the cross-correlation technique to extract the mean profile of K2-100's spectral lines by the following steps. First, we removed the telluric lines from the spectra using normalized spectra of the A0 star HD 71960 taken immediately before those of K2-100. In doing so, we took into account the difference in target airmass assuming a plane-parallel model of the atmosphere. We then cross-correlated each processed spectrum against a stellar template for K2-100.  For the template, we employed the theoretical synthetic model by \citet{Coelho2005} with $T_\mathrm{eff}=6125$ K,  $\log g=4.5$ dex, and [Fe/H] = 0. Finally, we aligned each resulting cross-correlation function (CCF) with the same velocity reference, taking into account the barycentric motion of the Earth for each frame. 

To determine the projected rotation velocity $v\sin i$ of K2-100, we co-added all the \emph{ out-of-transit} CCFs.  The dashed blue line in Fig. \ref{fig:ccf} plots the mean, normalized CCF profile that night. To compare this observed CCF profile with theoretical models, we generated a number of mock IRD spectra, starting from the synthetic spectrum of \citet{Coelho2005}.  We broadened the synthetic spectrum using the rotation plus macroturbulence line-broadening kernel with differing $v\sin i$ values (while the macroturbulence velocity was fixed at $\zeta=4.5$ \kms). We then calculated CCFs for these mock IRD spectra in the manner described above, and compared them with the observed CCF for K2-100. The best-fit model using $\chi^2$ minimization between the observed and model CCFs had $v\sin i=14.4 \pm 0.5$ \kms. The best-fit model is displayed as the solid red line in Fig. \ref{fig:ccf}. 

\begin{figure*}
	\includegraphics[width=17.5cm]{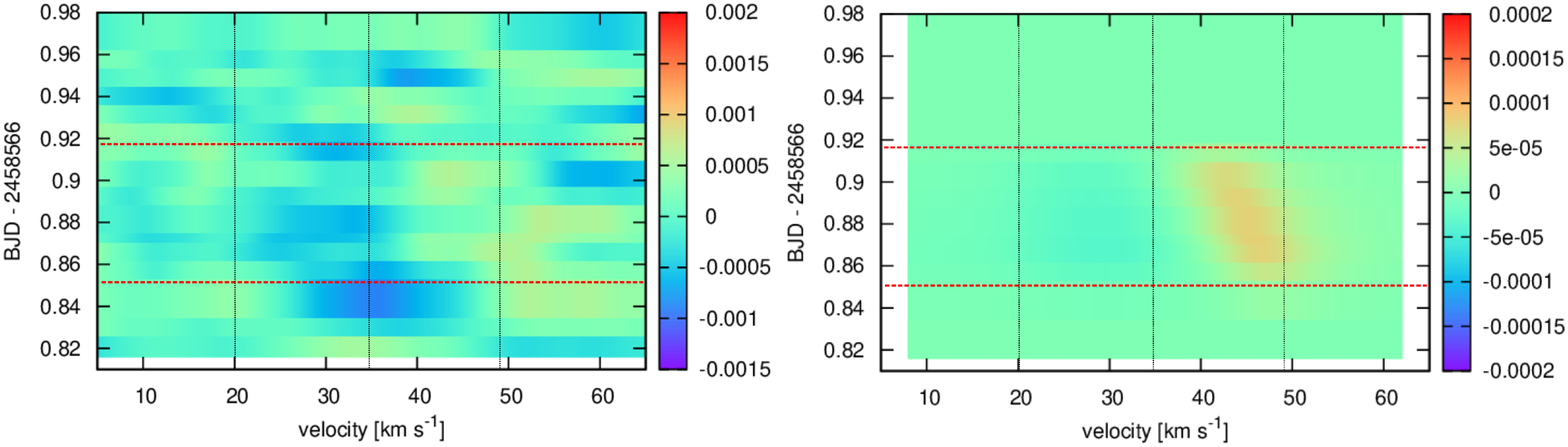}
    \caption{Left: Residual CCF vs. time of the spectrum of K2-100.  Right: Best-fit model with $\lambda = 108^{15}_{-14}$ deg but consistent with non-detection. In both panels, the transit ingress and egress times are indicated by the red horizontal lines. The approximate CCF center and its $\pm 14.4$ \kms{} ($\approx v\sin i$) are shown by the vertical black lines. Note that the colar-bar scales are different by about an order of magnitude between the two panels. }  
    \label{fig:dt}
\end{figure*}

We also attempted to visualise the line-profile variation during the transit, to search for the ``Doppler shadow" of K2-100b and estimate the projected stellar spin-orbit obliquity $\lambda$. We subtracted the mean {\it out-of-transit} CCF calculated above from individual CCFs, and inspected the variation in the residual CCF vs. time, as shown in Fig. \ref{fig:dt}. The noise level in the residual map is typically $0.0005-0.001$ of the normalized CCF. We also simulated the expected Doppler shadow, by creating the mock IRD spectra during the transit of K2-100b. As described in Hirano et al. (submitted) we simulated the IRD spectra during the transit for many different planet positions on the stellar disk and stellar rotational $v\sin i$, and computed the model CCF for each. 
Using these theoretical CCFs, we attempted to fit the observed residual CCF map for K2-100 with a few fitting parameters including $\lambda$ and $v\sin i$. Fig. \ref{fig:dt} shows the derived best-fit CCF residual map for K2-100's transit and best-fit model that is consistent with a non-detection. Unfortunately, the expected Doppler shadow of K2-100b is found to be by about an order of magnitude smaller than the observed noise in the residual CCF map, and we were unable to constrain the obliquity of K2-100 in this way. 

\bsp	
\label{lastpage}
\end{document}

%% file: newcommand_gen.tex
\newcommand{\twind}{$T_{w}$}

\newcommand{\mearthpergyr}{$M_{\oplus}$~Gyr$^{-1}$}

\newcommand{\hei}{He~I} 

\newcommand{\lyalph}{Lyman $\alpha$}

\newcommand{\galex}{\emph{Galex}}

\newcommand{\gaia}{\emph{Gaia}}
\newcommand{\ktwo}{\emph{K2}}

\newcommand{\kepler}{\emph{Kepler}}

\newcommand{\rosat}{\emph{ROSAT}}

\newcommand{\kms}{\mbox{km\,s$^{-1}$}}

\newcommand{\msun}{M$_{\odot}~$}

\newcommand{\rsun}{R$_{\odot}~$}
\newcommand{\lsun}{L$_{\odot}~$}

\newcommand{\mearth}{$M_\oplus$~}

\newcommand{\rearth}{$R_{\oplus}$}

\newcommand{\teff}{\ensuremath{T_{\rm eff}}}

\newcommand{\logg}{\ensuremath{\log{g}}}

